\documentclass[conference]{IEEEtran}
\IEEEoverridecommandlockouts
\usepackage{cite}
\usepackage{amsmath,amssymb,amsfonts}
\usepackage{algorithmic}
\usepackage{xcolor}
\usepackage{url} 
\usepackage{subfig}
\usepackage{float}
\usepackage{overpic}
\usepackage{graphicx}
\usepackage{textcomp}
\usepackage{soul, color}
\usepackage{multirow}
\usepackage[normalem]{ulem}
\def\BibTeX{{\rm B\kern-.05em{\sc i\kern-.025em b}\kern-.08em
    T\kern-.1667em\lower.7ex\hbox{E}\kern-.125emX}}

\newcommand{\eat}[1]{}

\usepackage{soul}
\usepackage{multicol}
\usepackage{latexsym}
\usepackage{amsfonts}
\usepackage{amsmath}
\usepackage{xcolor}
\usepackage{colortbl}
\usepackage{epsfig}
\usepackage{xspace}
\usepackage{graphicx}
\usepackage{subfloat}
\usepackage{paralist}
\usepackage{enumerate}
\usepackage[color,matrix,arrow,all]{xy}
\usepackage{comment}
\usepackage{booktabs}
\usepackage{balance}
\usepackage{stmaryrd}
\usepackage{pifont}
\usepackage{hhline}
\usepackage{listings}
\usepackage{array}
\usepackage{float}
\usepackage[flushleft]{threeparttable}
\usepackage{amsthm}
\usepackage{wrapfig}

\newcolumntype{L}[1]{>{\raggedright\let\newline\\\arraybackslash\hspace{0pt}}m{#1}}
\newcolumntype{C}[1]{>{\centering\let\newline\\\arraybackslash\hspace{0pt}}m{#1}}
\newcolumntype{R}[1]{>{\raggedleft\let\newline\\\arraybackslash\hspace{0pt}}m{#1}}

\usepackage{epsfig}
\usepackage{multirow}
\usepackage{url}

\usepackage{tikz}
\usetikzlibrary{shapes,snakes}
\usetikzlibrary{calc}

\usepackage[lined,boxed,vlined,ruled,linesnumbered]{algorithm2e}

\SetCommentSty{mycommfont}
\usepackage[export]{adjustbox}

\makeatletter
\newcommand{\removelatexerror}{\let\@latex@error\@gobble}
\makeatother

\NewDocumentCommand{\yuyu}{ mO{}}{\textcolor{blue}{\textsuperscript{\textit{Yuyu}}\textsf{\textbf{\small[#1]}}}}

\newcommand{\stab}{\vspace{1.2ex}\noindent}

\newcommand{\bi}{\begin{itemize}}
	\newcommand{\ei}{\end{itemize}}

\newcommand{\be}{\begin{enumerate}}
	\newcommand{\ee}{\end{enumerate}}
\newcommand{\beqn}{\begin{eqnarray*}}
	\newcommand{\eeqn}{\end{eqnarray*}}

\newcommand{\stitle}[1]{\vspace{1ex}\noindent{\bf #1}}

\newcommand{\etitle}[1]{\vspace{0.8ex}\noindent{\uline{\em #1}}}

\newcommand{\eg}{{\em e.g.,}\xspace}

\newcounter{ccc}

\DeclareMathAlphabet{\pazocal}{OMS}{zplm}{m}{n}

\newcommand{\eop}{\hspace*{\fill}\mbox{$\Box$}\vspace{1ex}}     

\newcounter{example}
\renewcommand{\theexample}{\arabic{example}}

\newcommand{\nthesection}{\arabic{section}}

\newcounter{definition}[section]
\renewcommand{\thedefinition}{\nthesection.\arabic{definition}}

\newcounter{alg}[section]
\renewcommand{\thealg}{\nthesection.\arabic{alg}}

\newcounter{arule}
\renewcommand{\thearule}{\arabic{arule}}

\newcounter{claim}
\renewcommand{\theclaim}{\arabic{claim}}

\makeatletter
\newcommand\figcaption{\def\@captype{figure}\caption}
\newcommand\tabcaption{\def\@captype{table}\caption}
\makeatother

\definecolor{shadecolor}{RGB}{200,200,200}

\definecolor{shadecolor1}{RGB}{230,230,230}

\definecolor{shadecolor1}{RGB}{255, 114, 118}

\tikzstyle{mybox} = [draw=black, fill=black!5, thick,
rectangle, rounded corners, inner sep=0pt, inner ysep=6pt]
\tikzstyle{fancytitle} =[fill=black, text=white]

\newcommand{\sys}{{\tt gCCTB}\xspace}
\newcommand{\ourmethod}{{\tt gCCTB}\xspace}
    
\begin{document}

\title{
GPU-Accelerated OLTP: An In-Depth Analysis of Concurrency Control Schemes 
}


\author{
    \IEEEauthorblockN{
        Zihan Sun\IEEEauthorrefmark{1}\IEEEauthorrefmark{2},
        Yuyu Luo\IEEEauthorrefmark{2} \thanks{Corresponding author: Yuyu Luo and Yong Zhang},
        Yong Zhang\IEEEauthorrefmark{1},
        Chao Li\IEEEauthorrefmark{1},
        Chunxiao Xing\IEEEauthorrefmark{1},
    }
    \IEEEauthorblockA{\IEEEauthorrefmark{1}Tsinghua University, China}
    \IEEEauthorblockA{\IEEEauthorrefmark{2}HKUST (GZ), China}
}

\maketitle

\begin{abstract}
Over the past decade, GPUs have demonstrated significant potential in accelerating Online Analytical Processing (OLAP) operations. However, there remains a substantial gap in their application to Online Transaction Processing (OLTP), as GPUs were traditionally considered unsuitable for such workloads. Despite this perception, the massive parallelism and high memory bandwidth of GPUs offer a unique opportunity to process thousands of transactions concurrently, making them promising candidates for OLTP acceleration.
Concurrency control schemes, which play a critical role in determining the performance of OLTP systems, may behave differently on GPUs due to their architectural differences from CPUs. This raises a key question: {\em How well do concurrency control schemes designed for CPUs adapt to GPU environments?}

To answer this, we present \ourmethod, the first testbed designed to evaluate concurrency control schemes on GPUs. We implement and benchmark eight CC schemes, including six classic CPU-oriented schemes and two designed specifically for GPUs, on both the YCSB and TPC-C benchmarks under varied contention levels and GPU configurations. Our findings reveal that GPU-optimized schemes do not consistently outperform CPU-oriented schemes, particularly under specific workloads and contention levels.
Moreover, GPU-specific parameters, such as the number of threads per warp and warps per block, significantly impact performance and require careful tuning. 
Finally, we find that conflict resolution overhead is a crucial factor influencing the performance of CPU-oriented schemes on GPUs, with optimistic concurrency control consistently minimizing this overhead and outperforming other CPU-oriented schemes across all workloads.
\end{abstract}


\section{Introduction}

Graphics Processing Units (GPUs) are specialized hardware initially designed for rendering and processing complex graphics on display devices. While individual GPU cores are less powerful than CPU cores, GPUs achieve remarkable performance through massive parallelism. By executing thousands of threads concurrently using the Single Instruction, Multiple Thread (SIMT) model~\cite{lindholm2008nvidia}, GPUs excel in parallel computing tasks. Moreover, GPUs offer significantly higher memory bandwidth than CPUs and have a distinct memory hierarchy. For example, GPUs provide programmer-accessible shared memory with bandwidth equivalent to that of L1 cache, making them particularly well-suited for data-intensive workloads.

Given these strengths, there has been increasing interest in leveraging GPUs to accelerate Database Management Systems (DBMS). GPUs have been successfully applied to a range of DBMS operations, including  
sorting~\cite{satish2009designing,satish2010fast,stehle2017memory,maltenberger2022evaluating}, join~\cite{rui2015join,sioulas2019hardware,lutz2020pump, rui2020efficient,lutz2022triton}, compaction~\cite{xu2020luda,homocomp} and index~\cite{awad2019engineering,awad2022gpu,zhang2015mega}, leading to significant performance gains.
Furthermore, several GPU-accelerated DBMSs~\cite{bress2014design,root2016mapd,lee2021art,heavydb} have been proposed for both research and commercial purposes. This growing interest in GPU-accelerated DBMS reflects the potential of GPUs to transform traditional database workloads by harnessing their parallelism and high memory bandwidth. 
To better understand these techniques, we categorize existing works into two scenarios: \textit{GPU-accelerated OLAP} and \textit{GPU-accelerated OLTP}.

\stitle{GPU-accelerated OLAP.}
Over the past decade, much of the research has focused on GPU-accelerated OLAP~\cite{codd1993providing}. 
This is because OLAP workloads consist of identical operations on massive data items. 
These identical operations have a simple control flow and seldom communicate with each other. 
These features make OLAP workloads well-suited to the data-parallel strengths of GPUs.
For instance, the \textbf{join} operator, which is a key operator in OLAP, consists of comparisons between tons of data items in several tables and can be well-parallelized and accelerated using GPUs~\cite{rui2015join,sioulas2019hardware,lutz2020pump, rui2020efficient,lutz2022triton}.
Another important operator, \textbf{sort}, which consists of comparisons and swaps between data items, is also suitable to be accelerated using GPUs~\cite{satish2009designing,satish2010fast,stehle2017memory,maltenberger2022evaluating}.
In addition, the \textbf{group by} operator~\cite{chen05groupby,morfonios06cure}, which aggregates data along different dimensions and hierarchies, plays an equally crucial role in OLAP queries. In recent years, there have also been a few emerging studies that explore group by computations on GPUs~\cite{luan2024groupby,wu2025groupby}.
These examples illustrate how GPUs can effectively harness OLAP's inherent data parallelism, leading to significant performance improvements.

\stitle{GPU-accelerated OLTP.}
In contrast to OLAP, OLTP~\cite{gray1992transaction} tasks require less computation but involve more complex control flows. Each transaction may follow a different execution path and state (\eg waiting for read or write operations to complete), which at first glance seems incompatible with the batch processing nature of GPUs.
However, the strength of GPUs lies in their ability to handle a large number of \textit{concurrent} threads despite weaker single-core performance compared to CPUs. This aligns well with the needs of transaction processing, where individual operations are relatively simple but demand high concurrency.
As a result, GPUs hold significant potential for accelerating OLTP workloads. When fully utilized, GPU hardware resources can provide a powerful new solution for OLTP systems. Inspired by this potential, 
several existing works~\cite{he2011high,boeschen2022gacco,wei2024} have already demonstrated the effectiveness of leveraging the massive parallelism of GPUs by processing transactions in batches.
Among the core components of an OLTP DBMS, \textit{Concurrency Control} (CC) schemes are particularly important. CC schemes manage the concurrent execution of transactions, ensuring consistency and isolation. These schemes are often the key bottleneck in transaction processing performance, making them critical targets for research in GPU-accelerated OLTP systems.

\stitle{Our Focus: Concurrency Control in GPU-accelerated OLTP.}
Although previous studies~\cite{he2011high,boeschen2022gacco,wei2024} have proposed new concurrency control schemes tailored for batch processing on GPUs, a critical research gap remains. Specifically, there is limited understanding of how well-established concurrency control schemes, originally designed and optimized for multi-core CPUs, perform when adapted to a GPU environment.

To address this gap, our research aims to answer the following key questions.

\stitle{Q1:} \textit{How do CPU-oriented concurrency control schemes perform on GPUs?} 
We will assess the throughput of these schemes under varying levels of contention and different proportions of write operations. We will also provide a detailed breakdown of transaction processing times, such as waiting and aborting, to help identify potential bottlenecks in these schemes when adapted to GPUs.

\stitle{Q2:} \textit{How do GPU-specific architectural parameters influence the performance of these schemes?}
On NVIDIA GPUs~\cite{cudacdoc}, threads are organized into warps, and several warps form a thread block, which differs from how threads are organized and scheduled on CPUs. We will investigate how the number of worker threads per warp and the number of warps per block impact the performance of concurrency control schemes.

\stitle{Q3:} \textit{How do different concurrency control design choices impact performance on GPUs?}
We will explore the impact of different design decisions—such as single-version vs. multi-version, no-wait vs. wait-die, and latch-based vs. latch-free—on performance. Understanding the performance implications of these design choices will provide insights into the best approaches for designing and optimizing GPU-specific concurrency control schemes.

\stitle{Contributions.}
Existing systems like DBx1000~\cite{staring14} and CCBench~\cite{tanabe2020ccbench} have been developed for benchmarking concurrency control schemes in CPU environments, but they lack support for GPU platforms, limiting their applicability in GPU-accelerated OLTP systems. Our work addresses this gap by introducing a testbed designed specifically for evaluating concurrency control schemes on GPUs.
%

\stab (1) \textbf{A New Testbed.} 
We presented \ourmethod with many reusable components for evaluating concurrency control schemes in the GPU-accelerated OLTP. \sys allows for easy and efficient testing of different CC schemes on GPUs using a variety of benchmarks under different configurations 
(Section~\ref{sec_testbed}).

\stab (2) \textbf{Comprehensive Evaluations.}
To answer the aforementioned three key questions, we implemented eight CC schemes in \sys, with six schemes designed initially for CPUs and two designed for GPUs. We then extensively evaluated existing CC schemes using YCSB and TPC-C benchmarks under various conditions
(Section~\ref{sec_exp}).

\stab (3) \textbf{New Experimental Findings.} Our experiments yielded several important insights (Section~\ref{sec_findings}):

\etitle{(3.1) Performance Comparison of GPU- and CPU-oriented Schemes.} 
Surprisingly, GPU-oriented concurrency control schemes do not always outperform their CPU-oriented counterparts. In scenarios characterized by low contention and a high proportion of read operations, CPU-oriented schemes often match or even surpass the performance of GPU-based schemes. For example, optimistic concurrency control schemes have demonstrated up to 1.5x better performance than GPU-oriented schemes, such as GaccO~\cite{boeschen2022gacco}, in medium contention and balanced read/write conditions. This suggests that under certain conditions, the parallelism of GPU-based methods does not fully leverage the strengths of the hardware.

However, the advantage of GPU-oriented schemes becomes apparent in scenarios with high write proportions and high contention. Under these conditions, GaccO can achieve a throughput that is 2-10 times higher than that of CPU-based methods, showcasing the GPU's capability to handle high concurrency effectively.

\etitle{(3.2) Impact of Warp and Thread Block Configurations.}
We observe that GPU kernel parameters, specifically the number of worker threads per warp (warp density, $wd$) and the number of warps per block (block size, $bs$), significantly affect transaction processing performance. However, larger values of these parameters do not always result in better performance.

Among these parameters, the choice of $wd$ has the most pronounced impact on performance. In scenarios with high transaction conflict, a lower warp density can achieve up to 10x the throughput of a higher warp density configuration. This indicates that, in high-contention scenarios, reducing the number of threads per warp may lead to better parallelization and reduced resource contention.

In terms of block size, the optimal configuration varies with the level of contention. When contention is low, the best performance is often observed when the block size is around half of the maximum block size. Conversely, under high contention, the block size should be close to the maximum value to maximize throughput.



\etitle{(3.3) Key Factor Influencing CPU-oriented Schemes on GPUs.}
The main factor affecting the performance of CPU-oriented schemes on GPUs is the overhead associated with conflict resolution under contention. Among these schemes, optimistic concurrency control stands out by minimizing this overhead, consistently outperforming other CPU-oriented schemes across various workloads.

To optimize these schemes, the focus should be on reducing conflict resolution overhead. Latch-free optimizations, in particular, significantly enhance performance, achieving up to 3x the throughput compared to latch-based implementations. In contrast, multi-version control and wait strategies provide only minimal performance improvements.

\section{Concurrency Control Schemes}\label{sec_bg}

\begin{table}[t!]
\caption{Categorization of Concurrency Control Schemes}
\label{tab:cc_schemes}
\resizebox{\columnwidth}{!}{
\begin{tabular}{c|c|c|c}
\hline
Category                          & Scheme           & Is Optimistic? & Platform \\ \hline
\multirow{2}{*}{Two-Phase Locking}       & \textbf{tpl\_nw} \cite{bernstein1981concurrency} & \ding{55}        & \multirow{6}{*}{CPU}       \\ \cline{2-3} 
& \textbf{tpl\_wd} \cite{bernstein1981concurrency} & \ding{55}         &       \\ \cline{1-3}
\multirow{4}{*}{Timestamp Ordering}      & \textbf{TO} \cite{bernstein1981concurrency}      & \ding{55}         &       \\ \cline{2-3} 
& \textbf{MVCC} \cite{bernstein1983multiversion}    & \ding{55}         &       \\ \cline{2-3} 
& \textbf{Silo} \cite{tu2013speedy}    & \ding{51}          &       \\ \cline{2-3} 
& \textbf{TicToc} \cite{yu2016tictoc}  & \ding{51}          &       \\ \hline \hline
\multirow{2}{*}{Conflict Graph Ordering} & \textbf{GPUTx} \cite{he2011high}   & \ding{55}         & \multirow{2}{*}{GPU}      \\ \cline{2-3} 
& \textbf{GaccO} \cite{boeschen2022gacco}   & \ding{55}         &       \\ \hline
\end{tabular}
}
\end{table}


Concurrency control schemes ensure the isolation property of transaction processing, meaning transactions access data concurrently as if they were executed independently~\cite{bernstein1979formal, silberschatz2011database, bernstein1981concurrency}. The core idea is to regulate the timing of read and write operations by transactions. Techniques such as locks~\cite{bernstein1979formal}, timestamps~\cite{bernstein1981concurrency}, and conflict graphs~\cite{silberschatz2011database} are used to enforce isolation, with conflict graph-based schemes being deterministic, while others are non-deterministic.
Another consideration is when to check if a transaction meets the isolation requirements. Pessimistic concurrency control schemes perform the check on every access to a data item, while optimistic schemes delay the check until the transaction is committed.

In this section, we introduce eight concurrency control schemes implemented in \ourmethod{}. 
Table~\ref{tab:cc_schemes} shows 
their characteristics.


\subsection{Two-Phase Locking}


Two-Phase Locking (2PL)~\cite{bernstein1979formal,eswaran1976notions} ensures correct transaction execution by requiring locks before any read/write, making it a pessimistic scheme. It supports shared (read) and exclusive (write) locks: shared locks can coexist, while exclusive locks conflict with all others. A transaction proceeds in two phases: a growing phase to acquire locks, and a shrinking phase where no new locks are allowed and all locks are released at commit or abort.

Deadlocks arise when transactions wait on each other’s locks. To address this, 2PL applies either prevention or detection. Prevention schemes include \textbf{tpl\_nw} (no-wait, abort if lock unavailable), \textbf{tpl\_wd} (wait-die, abort younger transaction), and \textbf{tpl\_ww} (wound-wait, abort older transaction’s competitor). Detection relies on wait-for graphs to identify cycles. We implemented \textbf{tpl
\_nw} and \textbf{tpl\_wd} in \ourmethod.
We did not implement the wound-wait scheme because the lack of system-level synchronization primitives on GPUs makes preemption difficult to implement.

\subsection{Timestamp Ordering}

\stitle{Basic Timestamp Ordering.}
Basic Timestamp Ordering (\textbf{TO})~\cite{bernstein1981concurrency} serializes transactions by timestamps. Each element maintains a Read Timestamp (RTS) and Write Timestamp (WTS). A transaction reading an element must have a timestamp newer than its WTS, and writing requires a timestamp newer than both RTS and WTS; otherwise, the transaction aborts and restarts. To avoid inconsistencies when updating WTS of uncommitted elements, TO requires acquiring a lock before modifying WTS and holding it until completion.

\stitle{Optimistic Concurrency Control (OCC).}
Optimistic Concurrency Control (\textbf{OCC})~\cite{kung1981optimistic} extends timestamp ordering by assuming conflicts are uncommon, thus postponing locking and validation until commit time. A transaction proceeds in three phases: during the \emph{read phase}, it executes in a private workspace without blocking; in the \emph{validation phase}, it checks whether its read and write sets remain valid, aborting if conflicts are detected; and in the \emph{write phase}, it applies updates to the database.

\ourmethod{} implements two widely used OCC variants: \textbf{Silo}~\cite{tu2013speedy} and \textbf{TicToc}~\cite{yu2016tictoc}. Both share the same three-phase structure: they lock all tuples in the write set (ordered by primary key) during validation, generate timestamps based on partial orders of conflicting transactions to avoid centralized allocation, and verify that no concurrent modifications occurred to the read set. Their main difference lies in timestamp management: TicToc maintains separate read and write timestamps for each record, while Silo updates only on writes and combines distributed timestamps with a global epoch ID for final ordering.

\stitle{Multi-version Concurrency Control (MVCC).}
In MVCC~\cite{bernstein1983multiversion}, each write creates a new version of a tuple with an associated timestamp, forming a version chain ordered by time. Each version is annotated with a start and end timestamp defining its validity interval; the most recent version has an infinite end time, and intervals of different versions do not overlap. A read operation selects the version whose validity interval contains its timestamp, ensuring correct visibility without blocking concurrent writes.

In \ourmethod{}, we implement a straightforward MVCC scheme derived from the multi-version adaptation of basic \textbf{TO}. This design preserves the simplicity of timestamp ordering while enabling non-blocking reads. More sophisticated MVCC variants, such as hybrids that integrate OCC (\eg Hekaton~\cite{diaconu2013hekaton,larson2011high}), are beyond our current scope and left for future work.

\subsection{Conflict Graph Ordering}
Conflict graph ordering determines transaction execution order by modeling conflicts as a directed acyclic graph (DAG)~\cite{silberschatz2011database}. Each vertex represents a transaction, and edges capture read–write or write–write dependencies on the same item. A topological sort of the DAG yields a serializable schedule that respects all conflicts. In our system, the graph is built and sorted during preprocessing, and we evaluate two GPU-based schemes leveraging this idea.

The first, \textbf{GPUTx}~\cite{he2011high}, assigns each transaction a rank derived from the conflict graph. Transactions with the same rank (a K-set) can run concurrently without additional control. For each data item, operations are ordered by transaction ID: the first operation gets rank 0, and subsequent conflicting operations increment the rank. A transaction’s final rank equals the maximum rank of all its operations.

The second, \textbf{GaccO}~\cite{boeschen2022gacco}, comes from a GPU-accelerated OLTP DBMS. It constructs a lock table during preprocessing, recording which transaction currently owns each data item. During execution, operations wait for the lock holder to finish and then update the table to release the item. Unlike \textbf{GPUTx}, \textbf{GaccO} treats all accesses—reads and writes alike—to the same item by different transactions as conflicts.

\section{A Testbed for GPU Concurrency Control}\label{sec_testbed}



In this section, we begin by discussing the design considerations for our testbed (Section~\ref{sub:design}). We then introduce the relevant hardware features of GPUs (Section~\ref{sub:background}), followed by a detailed explanation of the design and implementation of \sys (Section~\ref{sub:sysoverview}--\ref{sub:implement}).

\subsection{Design Consideration}
\label{sub:design}

The CPU and GPU architectures differ significantly in terms of thread models, memory hierarchies, and execution paradigms. Unlike CPUs, which typically handle a few threads with high single-threaded performance, GPUs employ thousands of lightweight threads organized into warps and thread blocks. These threads rely on specialized memory types and atomic operations that differ substantially from CPU-centric systems. As a result, programs designed for CPUs cannot be directly executed on GPUs. They must be re-implemented and recompiled using GPU-specific languages (e.g., CUDA) and adapted to fully exploit the GPU’s parallelism and memory hierarchy. These challenges highlight the need for a dedicated testbed that abstracts these complexities while enabling researchers to efficiently evaluate a variety of concurrency control (CC) schemes.

The primary goal of \ourmethod{} is to provide a flexible and efficient platform for testing various CC schemes on GPUs, with a broad range of benchmarks and test configurations. These configurations include table formats, index types, CC schemes, thread organizations, and other relevant parameters.

To achieve this goal, three key \textbf{challenges} must be addressed:

\begin{itemize}
    \item 
{\textbf{(C1)}}
How to effectively coordinate the resource allocation and transaction execution between CPUs and GPUs?

   \item{\textbf{(C2)}} How to offer researchers the flexibility to implement new benchmarks easily?

   \item {\textbf{(C3)}} How to assist researchers in integrating new CC schemes?
\end{itemize}

We will discuss how to tackle the above challenges in Section~\ref{sub:sysoverview}.

\begin{figure}[t!]
  \hspace*{-1em}
  \vspace{-0.5em}
\includegraphics[width=1.05\columnwidth]{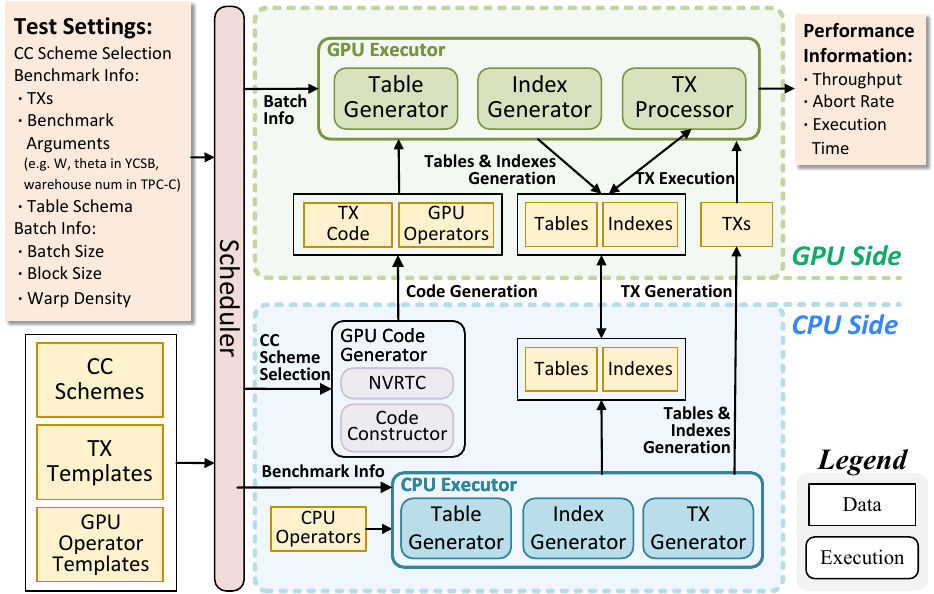}
  \vspace{-1.0em}
  \caption{The Architecture of the \sys Testbed.}
  \label{fig:testbed_arch}
  \vspace{-1.0em}
\end{figure}

\subsection{Backgound: GPU Architecture}
\label{sub:background}

GPUs exploit massive parallelism by organizing threads into blocks that are scheduled on Streaming Multiprocessors (SMs)~\cite{cudacdoc}. In the SIMT model~\cite{lindholm2008nvidia}, threads are grouped into warps of 32 that execute one instruction at a time, where branch divergence forces sequential execution of paths. Since Volta~\cite{voltadoc}, Independent Thread Scheduling has allowed each thread to maintain its own execution state for finer-grained control. CUDA also exposes a hierarchy of memory spaces~\cite{cudacdoc}: registers and local memory are private to each thread, shared memory is visible within a block and acts like a programmer-managed cache, and global and constant memory are accessible to all threads, with SM-private L1 caches and a device-wide L2 cache.

\subsection{An Overview of \sys}
\label{sub:sysoverview}

The architecture of \ourmethod{} is shown in Figure~\ref{fig:testbed_arch}. The testbed can be viewed as a compact database system consisting of both CPU and GPU components. The CPU handles scheduling and some computation tasks, while the GPU performs the majority of the computational work. Both components share the same table formats and maintain two copies of the database.

In the following, we describe how the CPU and GPU collaborate to address the challenges \textbf{(C1)-(C3)} discussed in Section~\ref{sub:design}.

For the first challenge \textbf{(C1)}, the CPU part coordinates the data generation, initialization, and transaction execution.
As a coordinator, it constructs tasks (such as data generation and transaction execution) according to test settings and then sends the corresponding data and code to the GPU part.
The GPU part conducts most of the computing tasks and then collects the runtime information for future performance analysis.
\ourmethod{} adopts a batch execution model \cite{boeschen2022gacco} to execute transactions of the same type on GPU in batches.
Each worker thread on GPU executes one transaction of the batch.

For the second challenge \textbf{(C2)},  an intermediate representation is designed for representing table formats and operations of \ourmethod{}.
The representation can be translated into definitions of data structures and executable operators on CPU or GPU.
Part of these operators are stored procedures while others whose functionalities depend on some dynamic configurations are compiled just in time.
With the combination of intermediate representation and just-in-time compilation, \ourmethod{} can easily generate well-optimized code and keep the flexibility of modifying testing configurations at run time.
A transaction is treated as a compiled-just-in-time procedure that is assembled with CC schemes and indexes specified by a configuration before testing.
Besides, all of the variables whose value can be determined before testing, such as a number of operations and memory addresses of data structures are also compiled into the transaction code as constants.
In this way, testers can implement and select different table formats, transactions, and data generation methods by writing and running multiple scripts without recompiling the whole testbed.


For the third challenge \textbf{(C3)},  \ourmethod{} provides an interface that is composed of a series of interface functions for integrating CC schemes and benchmarks into our testbed. 
These interface functions are abstracts of basic transaction operations. 
\verb|TxStart| and \verb|TxEnd| are called before and after the execution, respectively, to initialize and clean up the execution environment. 
\verb|Finalize| is called at the end of the execution procedure to collect statistics. 
CC schemes running on \ourmethod{} must implement this interface. 
A transaction uses these functions to manipulate data without knowing the details of CC schemes in advance.
To verify the correctness of CC scheme implementation, \ourmethod{} provides an event recording function to generate logs on GPUs and a verification program to analyze the log and check the correctness.
Moreover, \ourmethod{} also implements a few key techniques, such as locking and atomic operations on GPUs, which are discussed in detail below.

To illustrate how the components in Figure~\ref{fig:testbed_arch} work together, consider the execution of a sample transaction in \sys. A developer writes a transaction template that invokes \sys interfaces to express the logic, and separately specifies the read/write operations, table schemas, and index types. Based on this information, \sys generates tables and indexes on the CPU or GPU, constructs transaction batches on the CPU, and synchronizes data between the two sides. The GPU code generator integrates the template with a chosen concurrency control scheme to produce executable kernels, and the scheduler dispatches transaction batches to the GPU. During execution, the transaction logic accesses the declared indexes and calls \sys interfaces implemented by the selected concurrency control scheme, while performance metrics are collected for evaluation.

\subsection{Design Details}
\label{sub:details}

We first introduce the design of core components: code generation, measurement, and verification. Next, we discuss the design of tables and indexes in \ourmethod{}. Finally, we explore implementing two crucial techniques for CC schemes on GPUs: Spin Locks and Atomic Operations.


\stitle{Just-In-Time Code Generation.}
\ourmethod{} utilizes NVRTC~\cite{nvrtcdoc} to compile device code during runtime.
Device code is generated from predefined code templates with external macro definitions.
These macros are generated by dependencies of the templates dynamically and injected during JIT compilation.
To increase development efficiency, these templates have corresponding ``placeholder macros'' defined in the AOT part to give code editors correct highlighting and hints.

\stitle{Measurements and Verification.}
\ourmethod{} measures the execution time of different stages of transaction processing and accumulates it to help the researchers analyze the performance of CC schemes. 
The CPU part of \ourmethod{} times the duration of GPU batch execution and data transfer between CPU and GPU through the CUDA event mechanism.
The GPU collects the measurements in parallel to minimize the performance loss brought by time measurement: each thread measures its own execution time and stores it in its local space.
After all of the threads finish processing their workloads, they accumulate their local values to global measurements using atomic add operations.

Another critical aspect is verifying the CC schemes implemented on GPUs. \ourmethod{} uses a lightweight event system to collect key events (read, write, commit) during runtime on the GPU, which are then verified on the CPU post-execution. Each event is assigned a unique ID via an atomic counter, creating an ordered sequence. The verification program scans the sequence and constructs a conflict graph. If a loop is detected, it reports a bug and provides details, such as the timestamps of transactions in the loop, to help locate the issue. A macro controls the debugging function to minimize overhead during performance testing, and thanks to JIT compilation, it can be toggled dynamically at runtime.


\stitle{DBMS Building Blocks.}
A DBMS consists of several fundamental components. For the GPU side, we have adopted the following designs and implementations:

\begin{itemize}
    \item \textbf{Table}: A table on the GPU side adopts a row store, which is an array of tuples. The size of the table remains constant during GPU-side execution. All tables in a database are arranged consecutively in the GPU memory, sharing the same monotonically incremental ID as the primary key.
    \item  \textbf{Index}: Indexing on the GPU side is also a research field. Since indexes are not the focus of this paper, only sorted arrays~\cite{yen1990hash} are implemented as indexes on the GPU side. A column to be indexed is sorted and copied to the GPU memory together with the primary keys as the index. A query on the index is a binary search on the sorted array.
\end{itemize}






\stitle{Spin Lock.}
A spin lock~\cite{xu2016lock} is a lock that causes a thread trying to acquire it to wait in a loop while repeatedly checking whether the lock is available, which can be achieved by using \verb|atomicCAS| function provided by CUDA. Since GPUs lack a programmer-controllable thread scheduling mechanism, spin lock is the simplest way to implement a latch. The way of using the spin lock to protect a critical section on GPUs depends on the Independent Thread Scheduling (ITS) hardware feature, which enables a GPU to schedule each thread in a warp independently. 
On GPUs without ITS support, the critical section and release operation of a spin lock should be placed in the same loop as the compare-and-swap operation. Otherwise, the holder thread of a spin lock cannot reach the release code forever if another thread in the same warp is spinning to acquire the same lock because of the warp scheduling mechanism. 
Another key point is that a memory fence is necessary to be inserted after the release of a spin lock to confirm the memory order and visibility.  



\stitle{Atomic Operation.}
CC schemes often require critical sections that consist of a series of operations, which cannot be implemented with a single atomic function. Operations of the same critical section often read a 64-bit integer, do something else, and finally update the same integer. 
A natural way to implement critical sections is to use latches. However, this approach is inefficient since it requires extra memory access to the latch. Therefore, we adopt latch-free implementations of these critical sections. The main idea is to read the integer at the beginning and update it using \verb|atomicCAS| at the end of a loop. The only way to break the loop is if the integer has not been updated by other threads. In this way, the atomicity of a series of operations is reduced to the atomicity of a 64-bit integer.
Since the \verb|atomicCAS| function has a relaxed memory order, a memory fence is required to protect the sequentially consistent memory order, and the initial read of the integer should use a volatile pointer to read the most recent data.

\subsection{CC Schemes Migration and Implementation}
\label{sub:implement}

\begin{table*}[t!]
\small
\caption{Control Information Layouts of CPU-Oriented CC Schemes}
\label{tab:cc_schemes_layout}
\vspace{-1em}
\begin{center}
\begin{tabular}{l|l|l|l|l|l|l}
\hline
Scheme & \textbf{tpl\_nw} \cite{bernstein1981concurrency}                    & \textbf{tpl\_wd} \cite{bernstein1981concurrency}     & \textbf{TO} \cite{bernstein1981concurrency}                                                                           & \textbf{MVCC} \cite{bernstein1983multiversion}                                                                                                & \textbf{Silo} \cite{tu2013speedy}                                                           & \textbf{TicToc} \cite{yu2016tictoc}                                                                       \\ \hline
Size   & 8B                                                                                       & 8B           & 8B                                                                           & 16B                                                                                                 & 8B                                                             & 8B                                                                           \\ \hline
Layout & \begin{tabular}[c]{@{}l@{}}shared bit : 1\\ holder count : 31\\ holder : 31\end{tabular} & same as left & \begin{tabular}[c]{@{}l@{}}commit bit : 1\\ RTS : 31\\ WTS : 31\end{tabular} & \begin{tabular}[c]{@{}l@{}}commit bit : 1\\ RTS : 31\\ WTS : 31\\ version pointer : 64\end{tabular} & \begin{tabular}[c]{@{}l@{}}lock bit : 1\\ ts : 63\end{tabular} & \begin{tabular}[c]{@{}l@{}}lock bit : 1\\ delta : 15\\ WTS : 48\end{tabular} \\ \hline
\end{tabular}
\end{center}
\end{table*}

In this section, we introduce the migration and implementation of eight CC schemes on \ourmethod{}. All the concurrency control methods ensure the serializable isolation level, which represents the strongest form of transaction isolation.
Since \ourmethod{} adopts a batch execution model, the GPU memory consumption of a CC scheme can be determined in advance. Therefore, the CPU part of a CC scheme allocates and initializes the memory of its runtime information before the execution. 
It is common for CC schemes to load/store tuples of control information (timestamps, lock bit, other flag bits, etc.) atomically. According to our observation, these control information tuples can always be packed into 64 bits 
and be loaded/stored by a single memory operation. Thus, we utilize the C bit fields and the atomic operation mentioned above to achieve the latch-free load/store of the tuples to minimize the memory overhead, as exhibited in Table~\ref{tab:cc_schemes_layout}.
 
\subsubsection{Two-Phase Locking} \ourmethod{} implements two approaches of 2PL: \textbf{tpl\_nw} and \textbf{tpl\_wd}. 
There is no centralized lock manager in \ourmethod{}. Instead, each worker thread maintains its own lock information.
The information on a lock consists of a shared bit, lock holder bits, and holder count bits. 
Lock and unlock operations are achieved using the atomic operation mentioned above. We leave the wound-wait strategy unimplemented because it is hard for a GPU thread to terminate the execution of the transaction running on another thread in time.

\subsubsection{Timestamp Ordering} 
We discuss the timestamp ordering (TO) schemes implemented in \ourmethod{}.

\stitle{Basic TO.} The basic \textbf{TO} scheme acquires a new timestamp from an allocator in each iteration.
In our naive approach, it simply adds a 64-bit global variable atomically to allocate a new timestamp.
We adopt a 64-bit packed information structure, which consists of a commit bit, a 31-bit RTS, and a 31-bit WTS, and utilize the atomic operation. 
An implementation that supports longer timestamps has to use latches since the longer RTS, and WTS cannot be packed with a commit bit into a single 64-bit integer.

\stitle{MVCC.} The \textbf{MVCC} scheme implemented in \ourmethod{} is based on the basic \textbf{TO} scheme.
There is an array of the latest versions corresponding to all of the entries and another array of history versions.
Since the number of versions is the same as the number of write operations, the array of history versions can be pre-allocated and divided into local spaces maintained by worker threads.
A worker thread writing to an entry copies the corresponding version node into its local space and adds a new version to the global version array. 
For rolling back, the thread copies the previous version from its local space and returns it to the global array.
We try to pack the version pointer with timestamps into a 64-bit integer, but it result in a timestamp overflow.
Thus, we adopt the same timestamp format as \textbf{TO} and use separate version pointers.
The update of version pointers do not have to operate at the same time with the update of timestamps, so the aforementioned latch-free atomic operations can be utilized. Write operations of \textbf{MVCC} delay the update of version pointers to the commit time.

\stitle{OCC.} \ourmethod{} implements two OCC schemes, \textbf{Silo} and \textbf{TicToc}.
Both implementations utilize the atomic operation implementation described before to load entries and timestamps atomically.
They both assume that write operations are performed in ascending order of primary keys and that the write sets are not sorted when locking their items.
In the implementation of \textbf{Silo}, the timestamp information, including a lock bit and a 63-bit timestamp, is packed into a 64-bit integer.
Since transactions are processed in a batched manner on GPUs, there is no need to maintain the epoch described in the original \textbf{Silo} paper.
In the implementation of \textbf{TicToc}, timestamp information, including a lock bit, a 48-bit WTS, and a 15-bit delta, is packed into a 64-bit integer. 
Both schemes adopt the no-wait optimization mentioned in the original paper of \textbf{TicToc}.
This optimization makes a transaction abort instantly when waiting for a lock in the write phase.

\subsubsection{Conflict Graph Ordering} 
Two conflict graph ordering-based schemes, \textbf{GPUTx} and \textbf{GaccO}, are implemented on \ourmethod{}. 
In the preprocessing stage, implementations of both schemes first utilize the accessors mentioned above and indexes to gather all of the (\texttt{transaction id}, \texttt{primary key}) pairs, forming ``access tables''.
The preprocessing code that constructs the access tables is generated for workloads dynamically at runtime.
At the same time, they also count how many transactions each data item is accessed by to determine the boundary of each data item in the access table.
We adopt the thrust library~\cite{thrustdoc} for sorting and calculating prefix sums.
In \textbf{GaccO}, the preprocessing and execution of the same batch are assigned with the same CUDA stream.
Streams are synchronized before they access the same data structures on the GPU.
 
%

\section{Experiments}\label{sec_exp}

\subsection{Setup}

\stitle{Experimental Environment.} 
\ourmethod{} is implemented with C++ and CUDA library version 12.4.
The experiments are run on a server equipped with Intel$^\circledR$ Core\texttrademark \, i7-7700 CPU with 512 GB main memory and one NVIDIA RTX4090 GPU with 24 GB video memory.
The operating system is Ubuntu 20.04.6 LTS.

\stitle{Constraints.}
In this experiment, the executed transactions are constrained as follows:
\begin{itemize}
    \item All of the table and index data is stored in video memory in advance, and the updates and results generated by transactions are left in video memory. This avoids PCIe transfer overhead and allows the experiments to focus purely on GPU-side execution efficiency.
    \item The transactions only consist of read and write operations. There is no insert or delete operation, as insert and delete operations typically require complex support for dynamic memory management and index maintenance, which would introduce additional system-level concerns beyond concurrency control. Evaluating such operations is therefore left for future work.
    \item All of the read and write operations are determined before the experiment. This requirement follows the batch-processing nature of GPUs, as existing GPU-oriented concurrency control schemes~\cite{he2011high,boeschen2022gacco} also assume that the read/write sets are predetermined.
\end{itemize}

\stitle{Experimental Procedure.}
\ourmethod{} generates a fixed number of transaction requests and corresponding tables based on the selected benchmark and parameters, then transfers them to the GPU memory. \ourmethod{} creates and initializes instances of the CC scheme according to the selected scheme and parameters. It launches the CUDA kernel using the chosen startup parameters. On the GPU, each worker thread processes one transaction. If a transaction fails due to a conflict, the thread immediately restarts and re-executes the transaction until it commits successfully. During the initialization of the CC scheme and until all transactions are successfully executed, the testbed measures several selected metrics.

\stitle{CC Scheme Selection.}
Among the schemes described above, we select \textbf{GaccO}, \textbf{GPUTx}, \textbf{tpl\_nw}, \textbf{tpl\_wd}, \textbf{TO}, \textbf{MVCC}, \textbf{Silo} and \textbf{TicToc} as our research target. All the concurrency control methods ensure the serializable isolation level.

\stitle{Benchmark Selection.} 
We adopt the Yahoo! Cloud Serving Benchmark (YCSB) ~\cite{cooper2010benchmarking}, which models large-scale online services with transactions consisting of 16 single-tuple accesses following a Zipfian distribution.
We fix the table size in YCSB to $2^{20}*10$ rows, a number that fits within the limited GPU memory and is of the same order of magnitude as the row count used in the DBx1000 experiments~\cite{staring14}.  We fix the number of transactions in a batch to $2^{20}$.
There are two parameters $W, \theta$ controlling the proportion of write operations and contention level, respectively.
We construct three sets of YCSB benchmarks with different parameters, with the parameter selection being guided by the experimental design in DBx1000~\cite{staring14}:
\begin{itemize}
    \item Read Only (\textbf{RO}): $W=0, \theta=0$
    \item Medium Contention (\textbf{MC}): $W=0.1, \theta=0.6$
    \item High Contention (\textbf{HC}): $W=0.5, \theta=0.8$
\end{itemize}

Another benchmark adopted by us is TPC-C~\cite{tpccspec}, which simulates an order processing application.
Our evaluation focuses on the \verb|NewOrder| and \verb|Payment| transactions, which together comprise 88\% of the default workload mix. \verb|NewOrder| simulates order placement by updating stock and orderline records, while \verb|Payment| simulates customer payments, primarily contending on the warehouse table. To simplify analysis, we evaluate these transactions separately.

\stitle{Metric Selection.} 
We use the throughput and abort rate as the metrics. The throughput is determined by dividing the total number of transactions by the duration, measured in seconds, from the initialization of the concurrency control method to the successful completion of all transactions. The abort rate is defined as the ratio of aborts to successful transactions and may exceed one, as a single transaction can abort multiple times.
Moreover, we also measure the execution time of each stage of the transaction processing including preprocessing, CC manager working, waiting, index lookup, timestamp allocating, aborting and useful work. The aborting time is the sum of the execution time of the aborted transactions, subtracting the timestamp allocation duration.

\stitle{Launch Parameters.} 
There are two launch parameters that have significant impacts on the transaction processing performance in different ways:

\stab \textit{(1) Warp Density ($wd$)}: The number of working threads in a warp is $2^{wd}$, $wd \in [0,5]$. The non-working (idle) threads in a warp exit instantly right after the launch. 
    As $wd$ increases, both the level of contention and resource utilization decrease.

\stab \textit{(2) Block Size ($bs$)}: The number of warps in a thread block, $bs \in [1,32]$.
    As $bs$ increases, the parallelism and the resource utilization level of a single SM also increase. However, the average SM resources allocated to each thread of the same block decreased.

\subsection{Impact of Benchmark Parameters}


We first test the impact of transaction characteristics on GPU transaction processing performance. We test the schemes by separately controlling the YCSB contention level $\theta$ and the write proportion $W$. In both of these experiments, we fix the launch parameters as $wd=0, bs=32$. By fixing $wd=0$, we eliminate the influence of resource contention within a warp, such as warp divergence, on performance, as only one thread operates within the warp.

\begin{figure}[t!]
  \centering
  \vspace{-1em}
  \setlength{\abovecaptionskip}{0pt}
  \subfloat[Total Throughput]{
\includegraphics[width=0.49\linewidth]{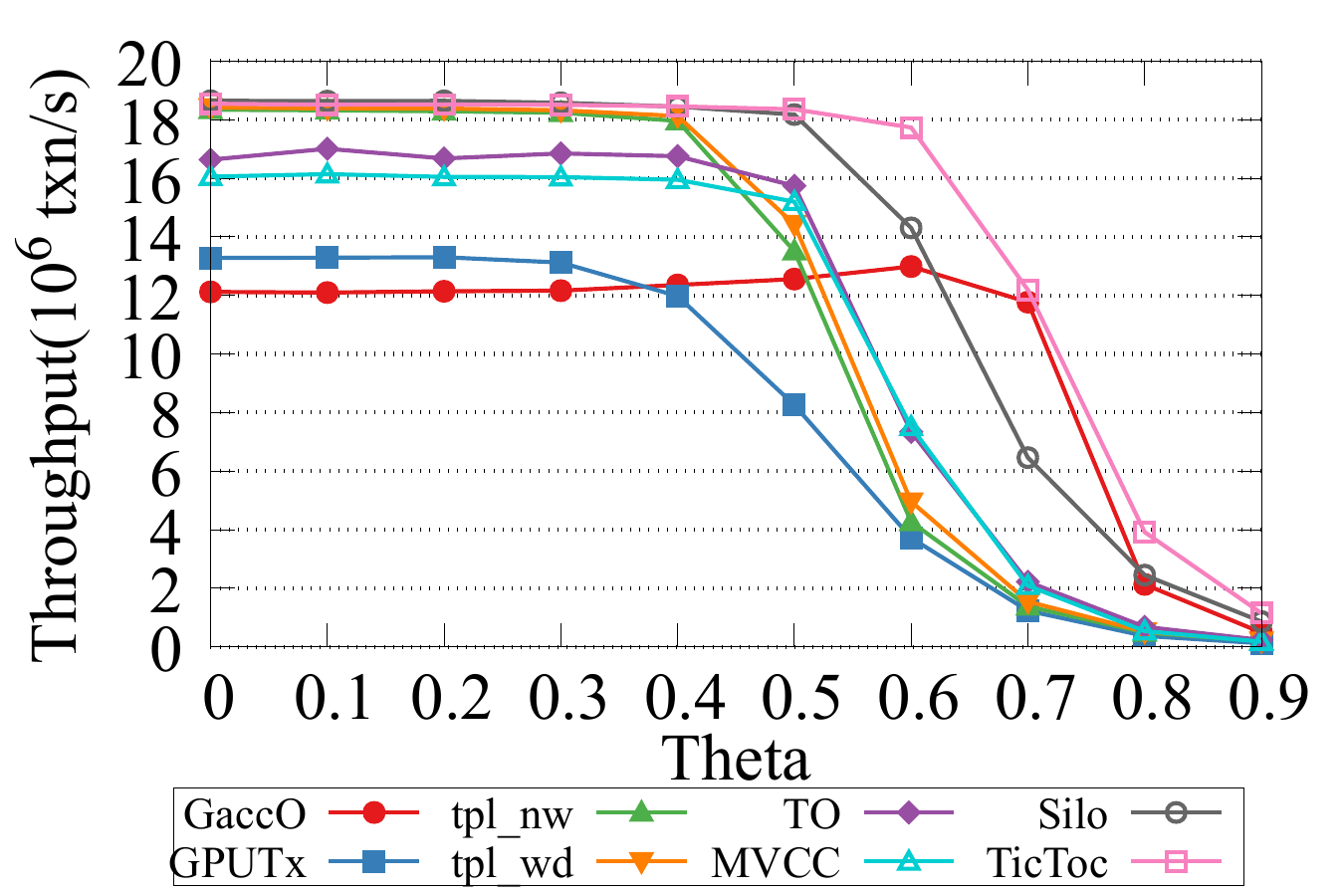}
  }
  \subfloat[Abort Rate]{
\includegraphics[width=0.49\linewidth]{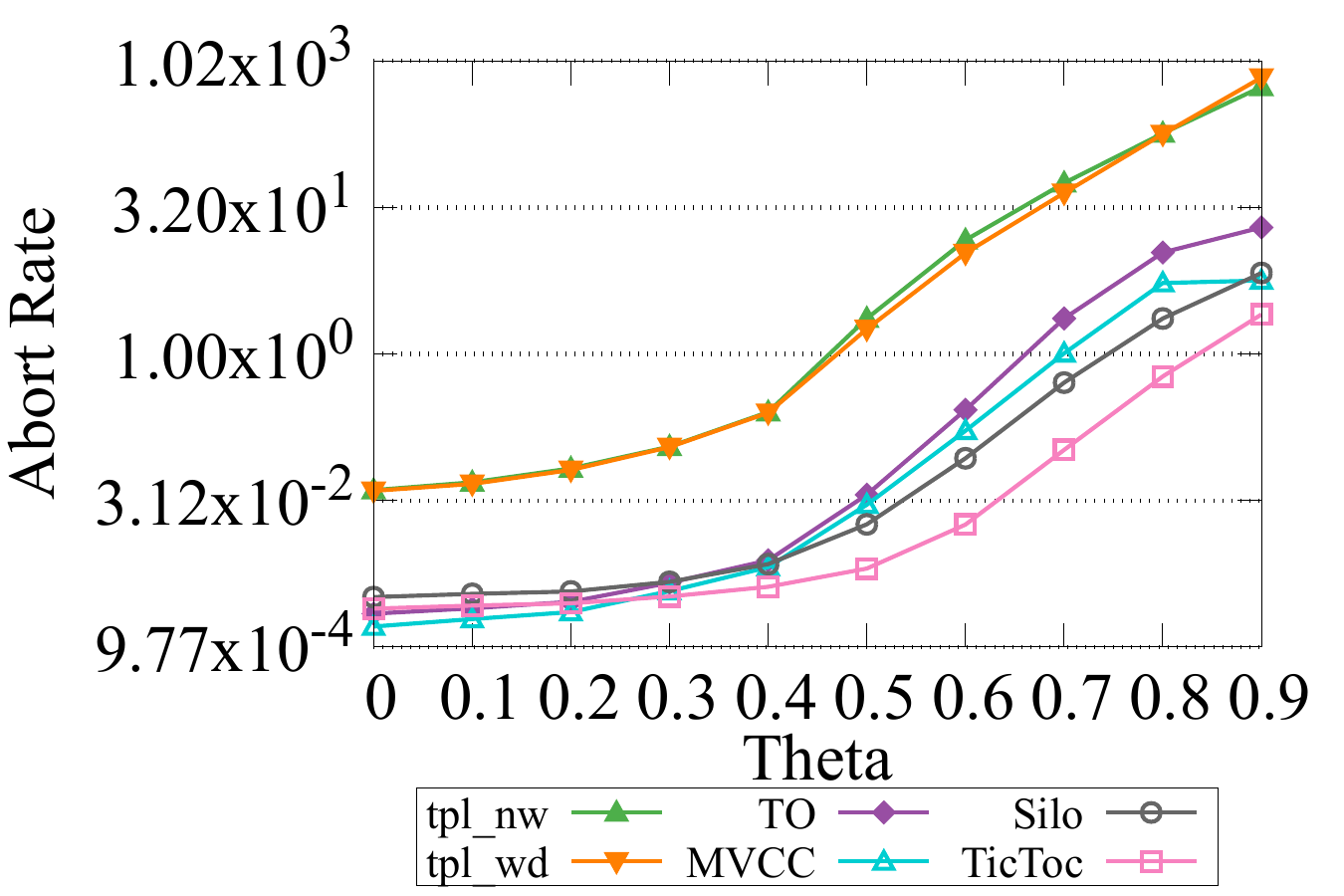}
  }
  \caption{YCSB. By Varying $\theta$ $W=0.1, wd=0, bs=32$}
  \label{fig:ycsb_difftheta}
\vspace{-2em}
\end{figure}

\begin{figure}[t!]
  \centering
  \setlength{\abovecaptionskip}{0pt}
  \subfloat[Total Throughput]{
  \includegraphics[width=0.49\linewidth]{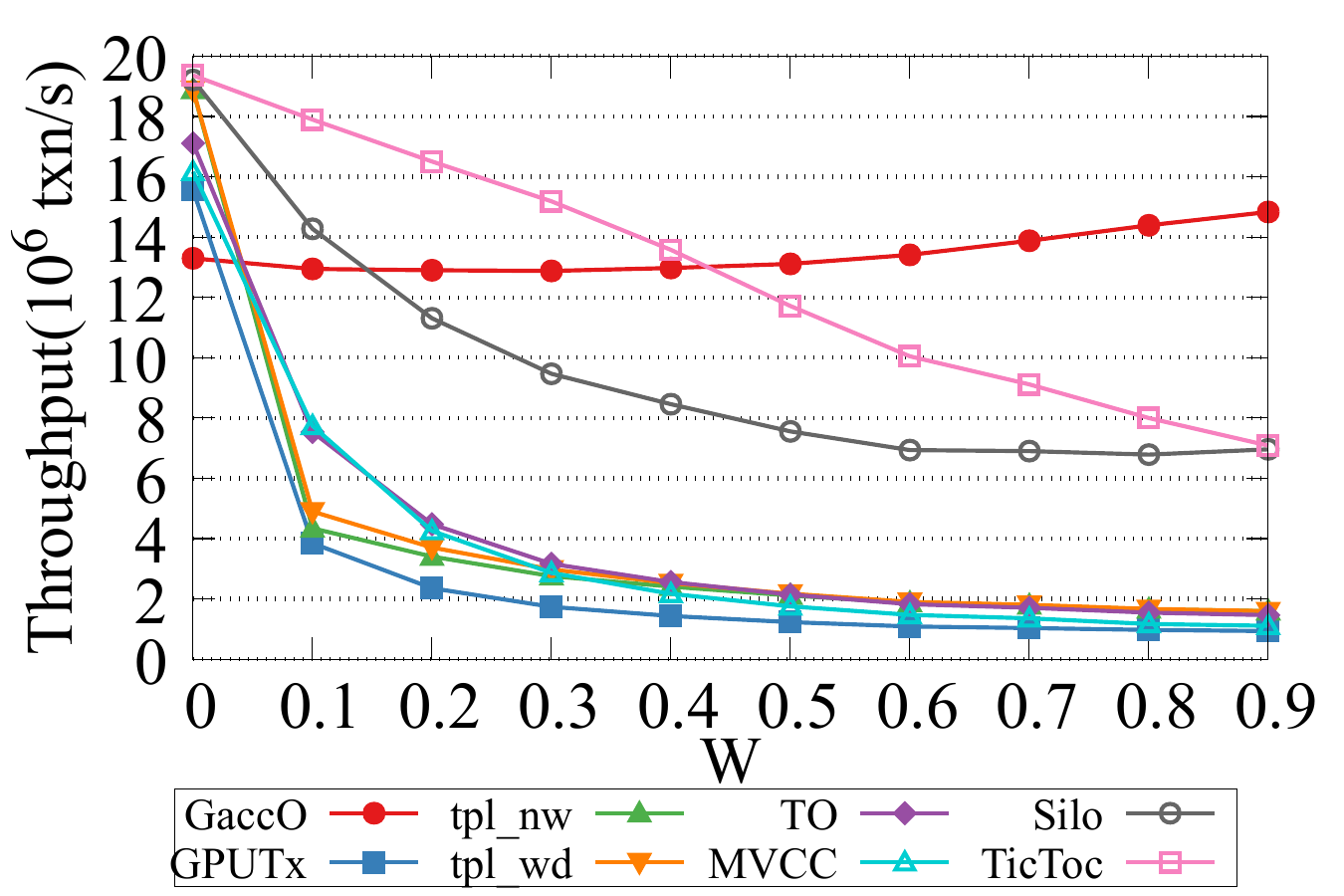}
  }
  \subfloat[Abort Rate]{
  \includegraphics[width=0.49\linewidth]{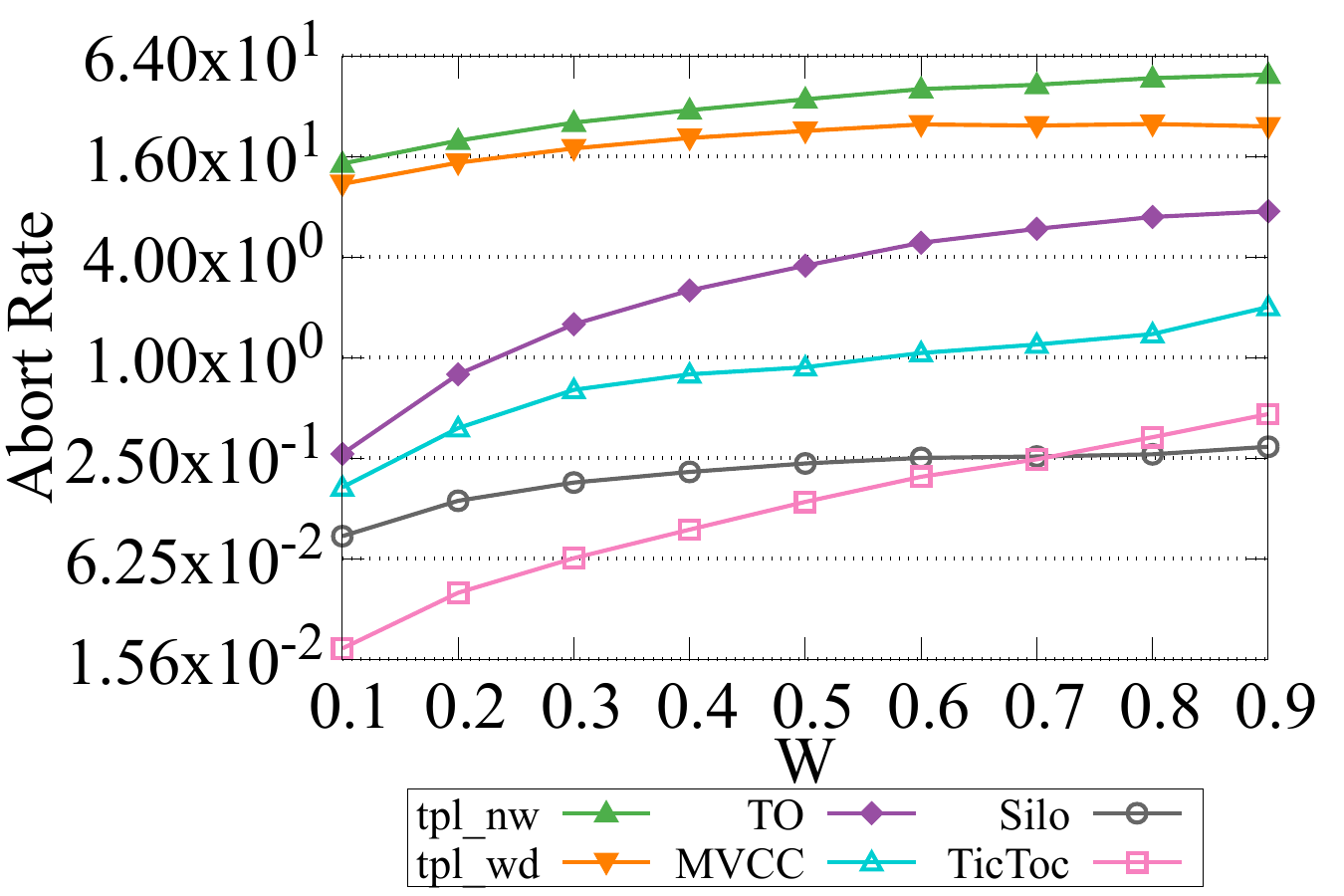}
  }
  \caption{YCSB. By Varying W $\theta=0.6, wd=0, bs=32$}
  \label{fig:ycsb_diffw}
\vspace{-2.em}
\end{figure}

\stitle{Exp-1: Impact of Contention Level (By Varying $\theta$).}
We fix the write proportion $W$ to 0.1 and vary the contention level $\theta$ from $0$ to $0.9$.

We make the following observations based on Figure~\ref{fig:ycsb_difftheta}.

\begin{enumerate}
   
\item The throughput suffers a sharp decrease as $\theta$ increases. The increase in abort rate is probably the main reason for the performance decrease for CC schemes other than conflict graph-based schemes. 

\item Especially for two-phase locking schemes, the abort rate is much higher than other schemes and reaches near 1024.

\item Although the abort rate is the highest, the performance of 2PL schemes is at the top level until $W=0.5$. When $W\geq0.5$, \textbf{MVCC} and \textbf{TO} beat 2PL.

\item To our surprise, OCC schemes always perform better than other non-deterministic schemes, even in high-conflict situations.

\end{enumerate}


\stitle{Exp-2: Impact of Write Proportion (By Varying $W$).}
We fix the contention level $\theta$ to $0.6$ and vary the write proportion $W$ from $0$ to $0.9$.

We make the following observations based on Figure~\ref{fig:ycsb_diffw}.

\begin{enumerate}
    
\item  The performance of CC schemes, except for \textbf{GaccO}, decreases when transactions start to contain write operations. 

\item  The throughput of \textbf{GaccO} even grows slightly with the increase of $W$ and beats all the other schemes after $W=0.5$. 

\item  When $W$ increases, the performance decrease of optimistic CC schemes is less dramatic than that of other schemes. 

\item  Schemes other than OCC and \textbf{GaccO} perform similarly with each other, and the performance of \textbf{TO} and \textbf{MVCC} is slightly higher than others.

\item The changes of throughput and abort rate when $W>0$ are more slightly than they vary with $\theta$. 

\end{enumerate}



\stitle{\underline{Insight 1:}} 
GPU-oriented schemes cannot beat CPU-oriented schemes in all cases.
When both $W$ and $\theta$ are low, most of the CPU-oriented schemes outperform GPU-oriented schemes.
Moreover, OCC schemes can still beat GPU-oriented schemes when $W$ grows in a medium-contention scenario.
This is because GPU-oriented schemes take a lot of time to pre-process to determine the execution order of transactions,
even if there are not as many conflicts to resolve.

\begin{figure}[t!]
  \centering
  \vspace{-1em}
  \includegraphics[width=\linewidth, height=0.2\textheight]{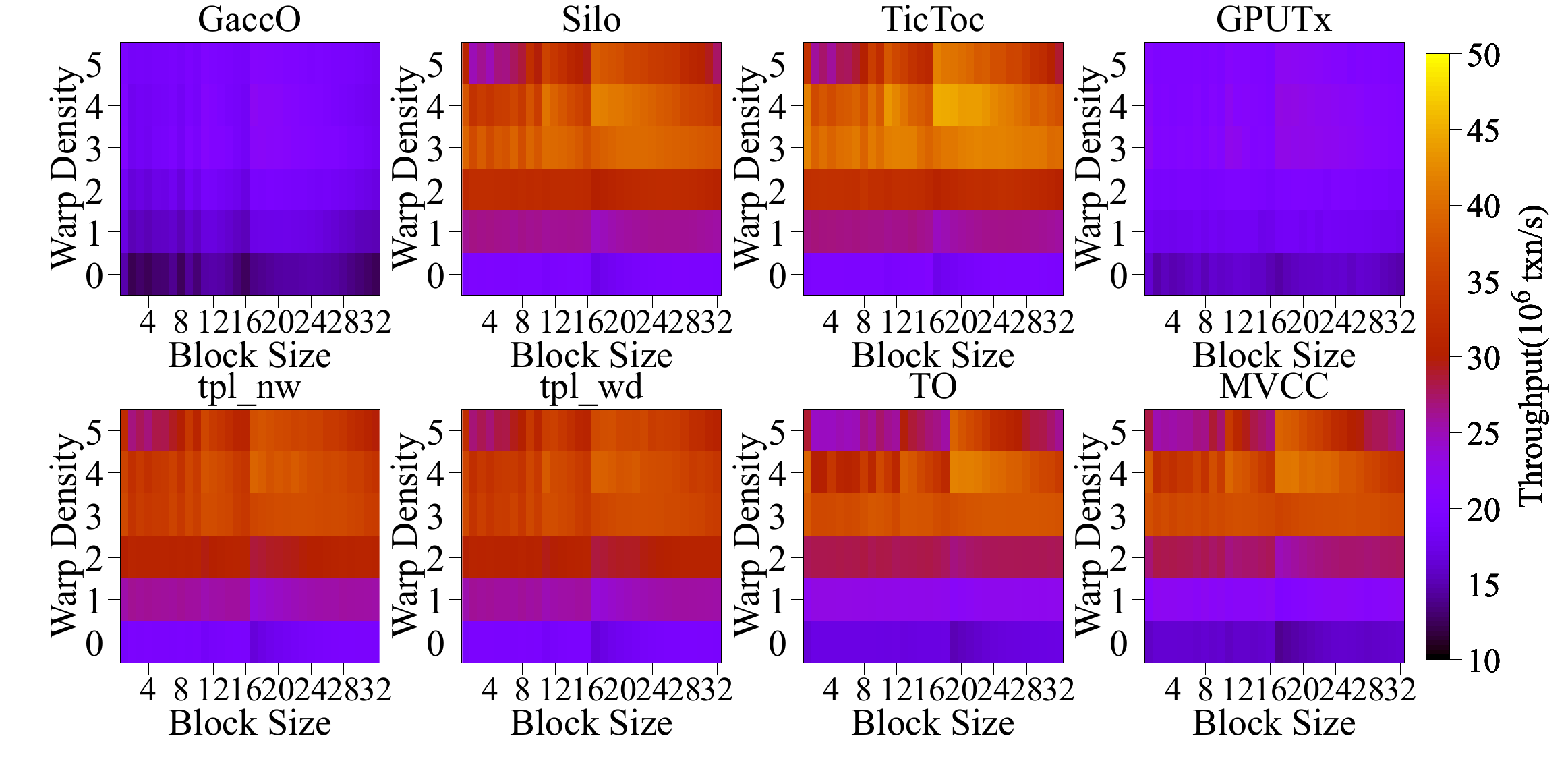}
  \vspace{-2em}
  \caption{Heatmap of YCSB-RO}
  \label{fig:ycsb_ro_heat}
\vspace{-1em}
\end{figure}

\subsection{Impact of Launch Parameters}

We test the impact of launch parameters block size $bs$ and warp density $wd$ on GPU transaction processing performance. We employ a grid search approach to observe the performance impact when both parameters vary simultaneously. Subsequently, we fix one parameter while varying the other to examine their individual effects on performance. These experiments are conducted on three preset YCSB benchmarks.


\stitle{Exp-3: Impact of Launch Parameters on YCSB-RO.}
Figure~\ref{fig:ycsb_ro_heat} shows the throughput distributions of schemes under different $bs$ and $wd$ on YCSB-RO.
The heatmaps are colored using the same data range.
An interesting phenomenon is that the performance of these schemes does not peak at the maximum $wd$, but at $wd=4$.
Although an increase in $wd$ can make more threads in a warp process transactions, it also exacerbates the uncoalesced memory access problem,
which causes threads to wait for one another to read from global memory.

\begin{figure}[t!]
    \centering
    \includegraphics[width=\linewidth]{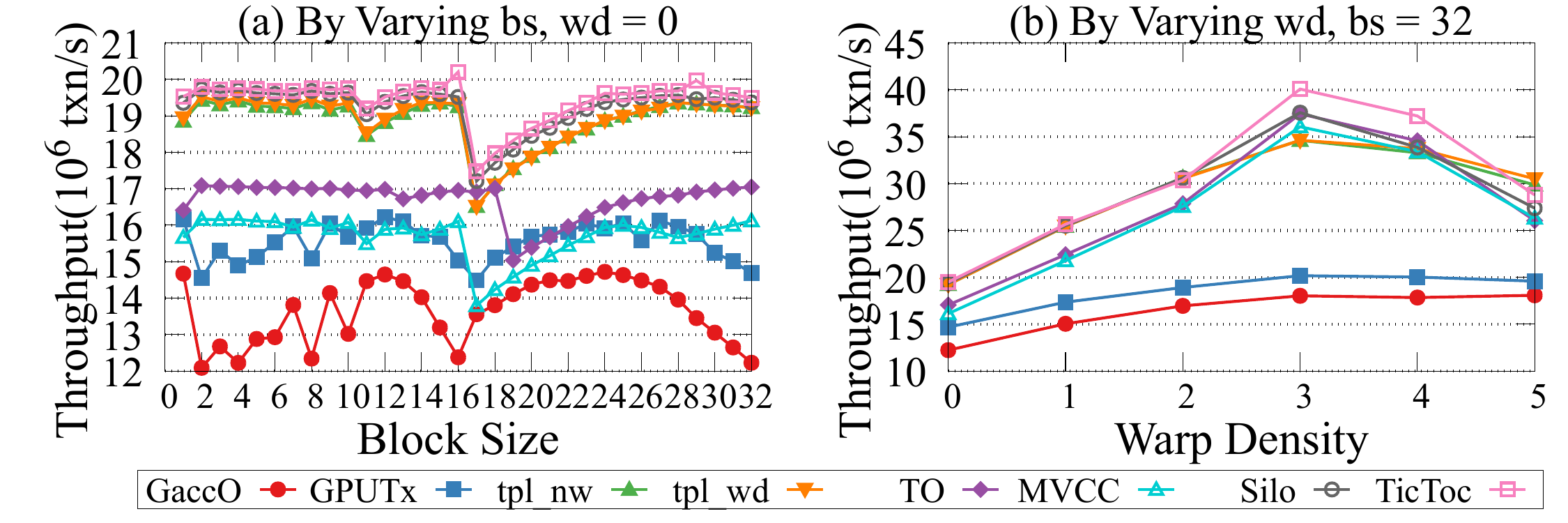}
    \vspace{-2em}
    \caption{YCSB-RO}
    \label{fig:ycsb_ro}
    \vspace{-2em}
\end{figure}

\begin{figure}[t!]
  \centering
  \vspace{1.5em}
  \includegraphics[width=\linewidth]{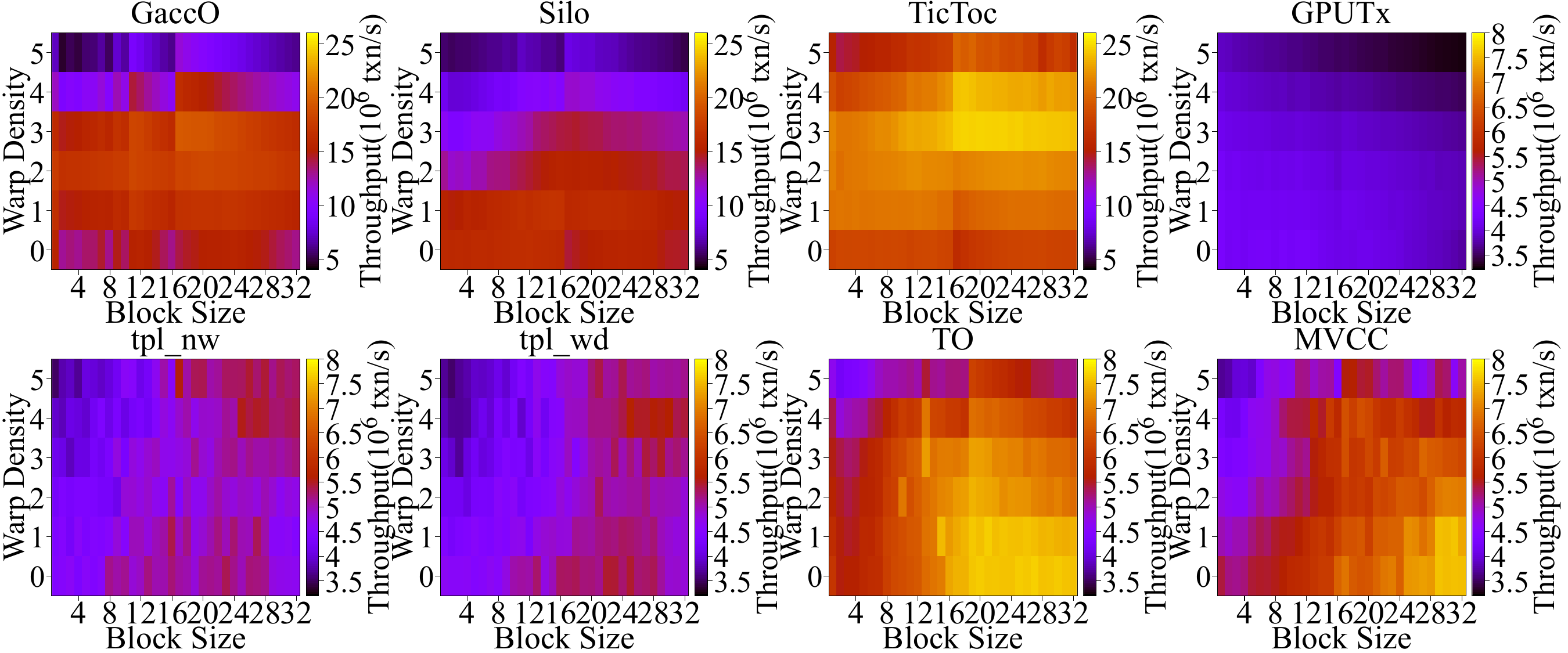}
    \vspace{-1em}
  \caption{Heatmap of YCSB-MC}
  \label{fig:ycsb_mc_heat}
\vspace{-2em}
\end{figure}

Figure~\ref{fig:ycsb_ro} shows how the throughput of schemes changes with $bs$ and $wd$ on YCSB-RO. We make the following observations:

\begin{enumerate}

\item When processing read-only transactions, all the schemes reach a throughput over $10^7$.

\item 2PL and OCC schemes exhibit the best performance in subplot (a), and nearly all of the non-deterministic schemes outperform conflict graph ordering schemes in both subplots due to the preprocessing overhead of conflict graph ordering schemes.

\item On YCSB-RO, \textbf{GPUTx} executes all the transactions in a single batch, and it does not introduce additional overhead when executing, so it performs better than \textbf{GaccO}.

\item Although \textbf{MVCC} operates the same as \textbf{TO}, it still gets a lower performance. We attribute it to the cache efficiency: timestamps of \textbf{TO} are densely arranged in memory while timestamps and version pointers of \textbf{MVCC} are arranged at intervals. 

\item It can be observed from (a) that the line shapes of non-deterministic schemes are similar. This is due to the similarity in their hardware occupancy.

\item In subplot (b), the throughput of all schemes grows with $wd$ when $wd\leq3$. However, the throughput of conflict graph ordering schemes remains stable, while other schemes face performance degradation. 
 
\end{enumerate}


\stitle{Exp-4: Impact of Launch Parameters on YCSB-MC.} Figure~\ref{fig:ycsb_mc_heat} shows the throughput distributions of schemes under different $bs$ and $wd$ on YCSB-MC.
The heatmaps of \textbf{GaccO}, \textbf{Silo}, and \textbf{TicToc} are colored using one data range, while others are colored using another data range.
Except for \textbf{TicToc} and \textbf{GPUTx}, the brighter areas of the heatmaps appear in the lower right, and the dark areas appear in the upper left.
Since threads in the same warp execute the same instruction at the same time, threads in a warp with higher $wd$ are easier to conflict with each other,
which causes higher abort overhead. 
A thread block is assigned to a single streaming multiprocessor (SM) and contains, at most, 1024 threads.
When $bs$ is fixed, a decrease in $wd$ means a decrease in the number of active threads in the thread block, which causes the underutilization of the SM resource.
Therefore, as $wd$ decreases, increasing $bs$ can help improve performance.

\begin{figure}[t!]
    \centering
    \includegraphics[width=\linewidth]{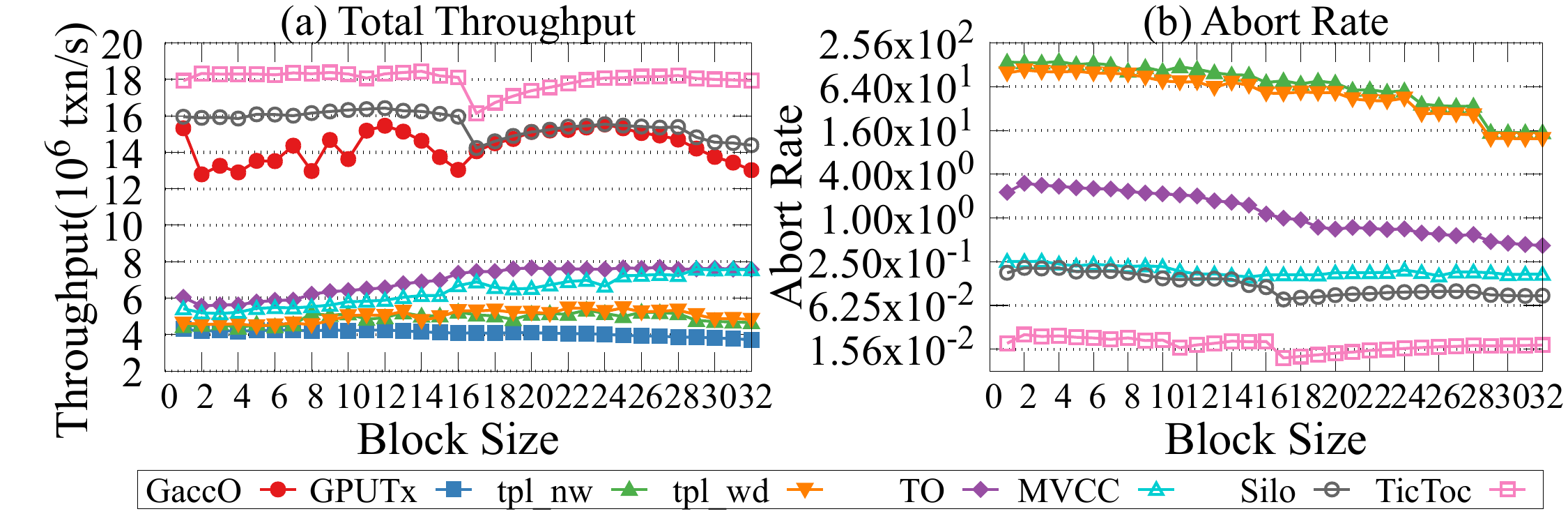}
    \vspace{-2em}
    \caption{YCSB-MC. By Varying $bs$, $wd=0$}
    \label{fig:ycsb_mc_diffbs}
\end{figure}

\begin{figure}
    \centering
    \vspace{-1em}
    \includegraphics[width=\linewidth]{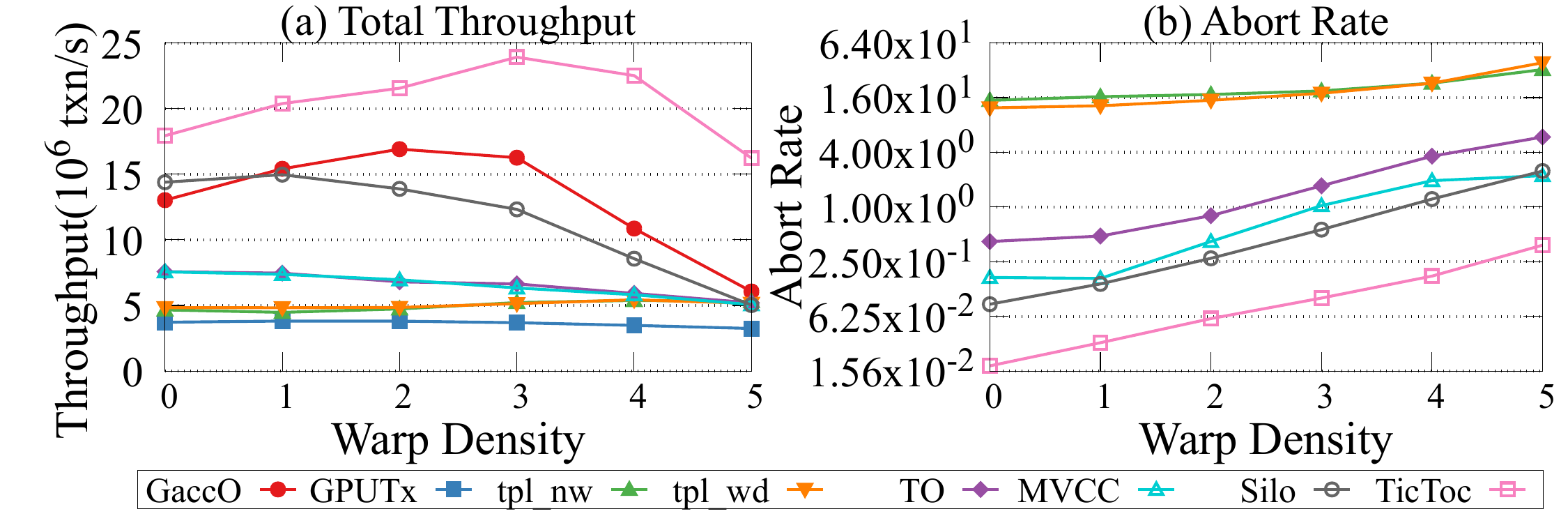}
    \vspace{-2em}
    \caption{YCSB-MC. By Varying $wd$, $bs=32$}
    \label{fig:ycsb_mc_diffwd}
    \vspace{-2em}
\end{figure}

\begin{figure}[t!]
  \centering
  \vspace{1.5em}
  \includegraphics[width=\linewidth]{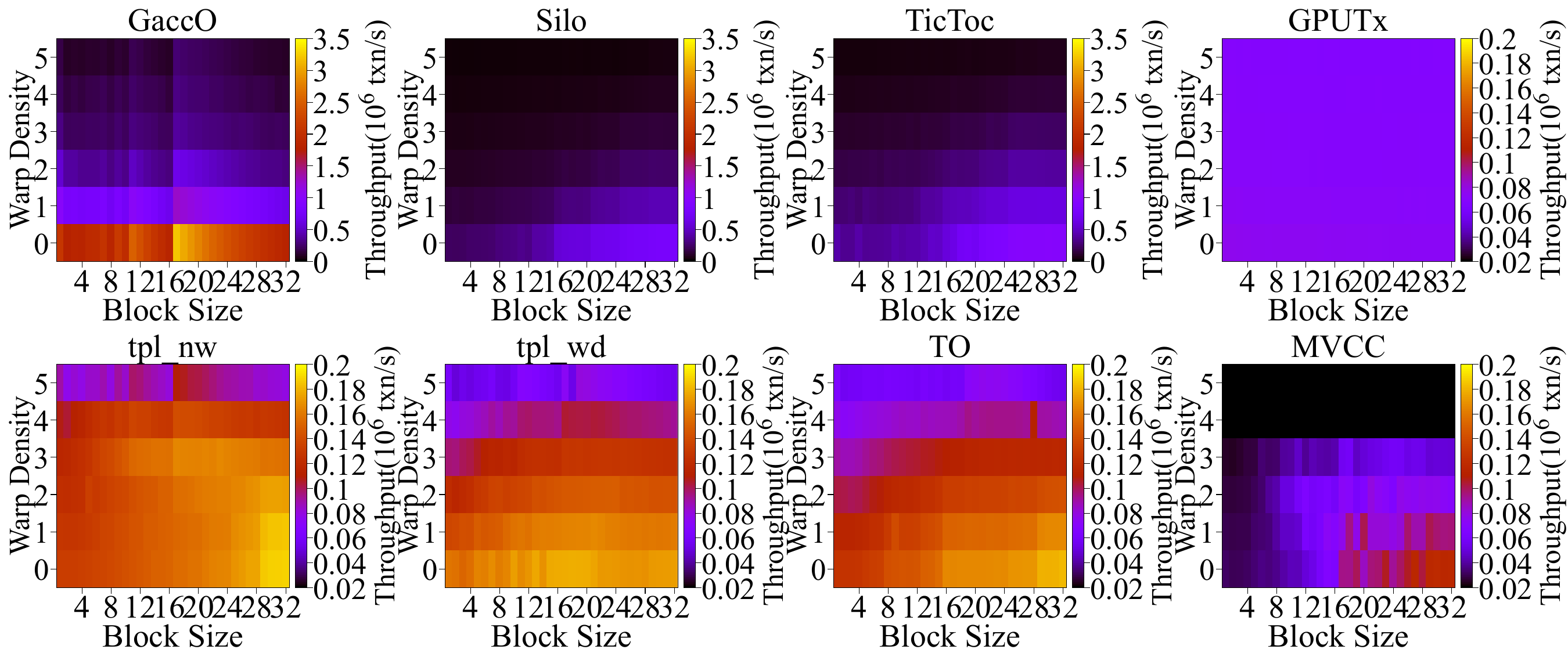}
  \vspace{-1em}
  \caption{Heatmap of YCSB-HC}
  \label{fig:ycsb_hc_heat}
\vspace{-2em}
\end{figure}

Figures~\ref{fig:ycsb_mc_diffbs} and~\ref{fig:ycsb_mc_diffwd} show how the throughput and abort rate of schemes change with $bs$ and $wd$ on YCSB-MC, respectively. We make the following observations from them:

\begin{enumerate}
    
\item The throughput of \textbf{GaccO} and OCC schemes is still above $10^7$, while the throughput of others is around $5 * 10^6$.

\item In both figures, the performance of \textbf{TicToc} is much higher than other schemes, and its abort rate is the lowest.

\item The performance of \textbf{TO} and \textbf{MVCC} is between OCC and 2PL, and the gaps between \textbf{TO} and \textbf{MVCC}, \textbf{tpl\_nw} and \textbf{tpl\_wd} are relatively small.

\item As write operations increase, the performance of \textbf{GPUTx} drops due to its large scheduling granularity.

\item The abort rate of \textbf{Silo} is slightly lower than that of \textbf{MVCC} and much higher than that of \textbf{TicToc}; this is because there is no separate read and write timestamps in \textbf{Silo}.

\item Although the abort rate of \textbf{MVCC} is much lower than \textbf{TO}, their throughput is close, which is because of the overhead of scanning the version chain.

\item According to Figure \ref{fig:ycsb_mc_diffwd}, the abort rate of schemes except 2PL grows exponentially with $wd$, which is consistent with the monotonic increasing throughput of \textbf{tpl\_nw} and \textbf{tpl\_wd} in subplot (a).
 
\end{enumerate}


\stitle{Exp-5: Impact of Launch Parameters on YCSB-HC.} Figure~\ref{fig:ycsb_hc_heat} shows the throughput distributions of schemes under different $bs$ and $wd$ on YCSB-HC.
The rules of coloring are the same as Figure~\ref{fig:ycsb_mc_heat}.
All the schemes perform better when $wd$ decreases, which proves the insight given in Figure~\ref{fig:ycsb_mc_heat}.

\begin{figure}[t!]
    \centering
    \includegraphics[width=\linewidth]{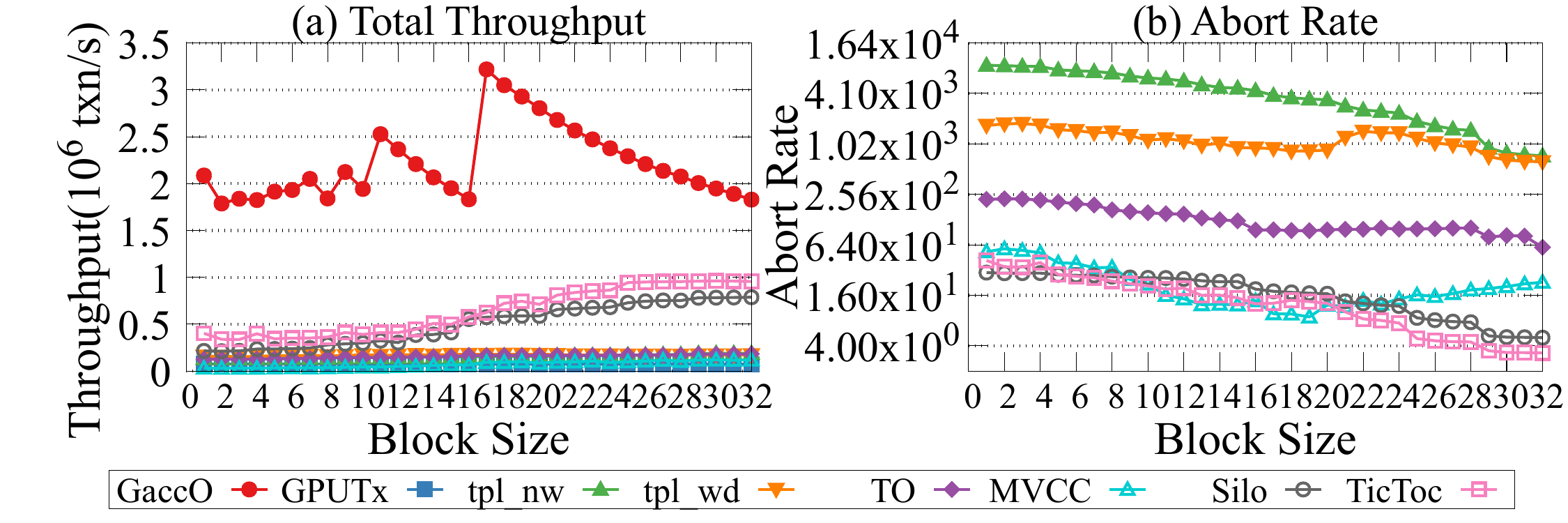}
    \vspace{-2em}
    \caption{YCSB-HC. By Varying $bs$, $wd=0$}
    \label{fig:ycsb_hc_diffbs}
    \vspace{-1em}
\end{figure}

\begin{figure}
    \centering    \includegraphics[width=\linewidth]{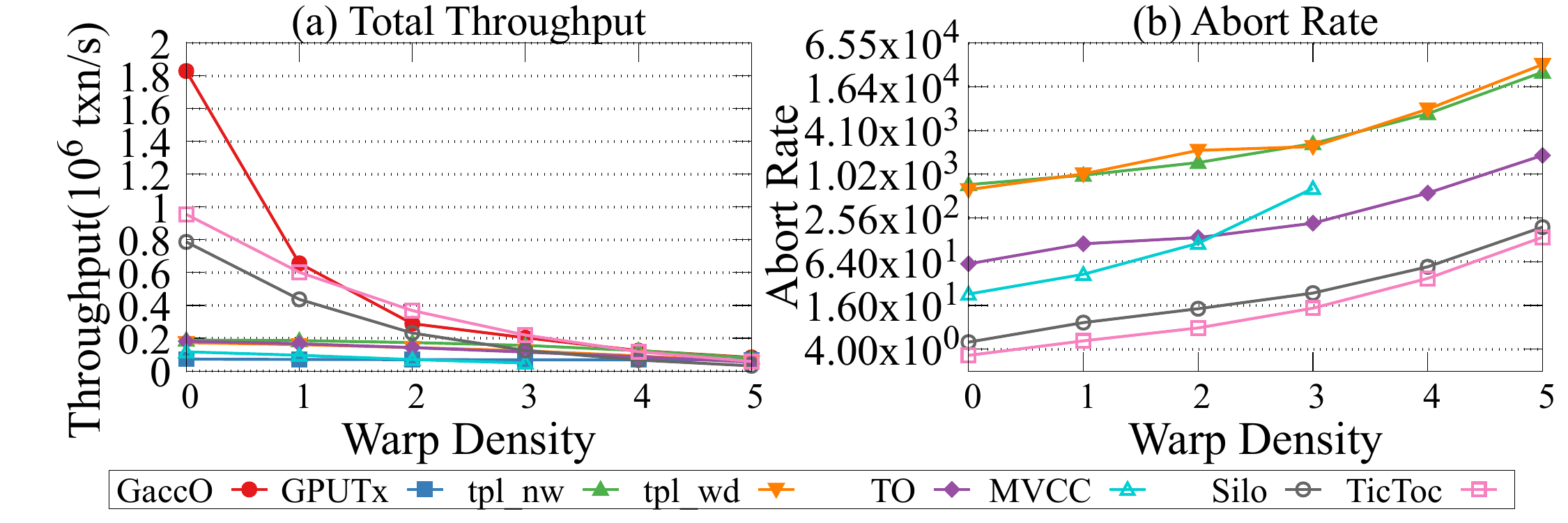}
    \vspace{-2em}
    \caption{YCSB-HC. By Varying $wd$, $bs=32$}
    \label{fig:ycsb_hc_diffwd}
\vspace{-2em}
\end{figure}

Figures~\ref{fig:ycsb_hc_diffbs} and~\ref{fig:ycsb_hc_diffwd} show how the throughput and abort rate of schemes change with $bs$ and $wd$ on YCSB-HC, respectively. We make the following observations from them:

\begin{enumerate}
    
\item As the contention and write proportion increase, the throughput of CC schemes except \textbf{GaccO} fails to reach $10^6$.

\item The throughput of \textbf{GaccO} is also an order of magnitude lower than when it was on YCSB-MC.

\item The throughput of OCC schemes is still higher than other schemes except \textbf{GaccO} and grows with $bs$.

\item The performance of other schemes does not change much with $bs$, although their abort rates decrease as $bs$ increases.

\item As shown in Figure \ref{fig:ycsb_hc_diffwd}, $wd$ has a huge impact on performance.

\item The abort rate of all the non-deterministic grows exponentially with $wd$.

\item The abort rate of \textbf{MVCC} grows even faster than \textbf{TO} and exceeds it when $wd=3$. Due to the high abort rate, \textbf{MVCC} faces timestamp overflow when $wd>3$ and fails to get a valid performance.
 
\end{enumerate}


\begin{figure*}[t!]
  \centering
  \begin{minipage}[b]{0.495\textwidth}
    \centering
    \vspace{-1em}
    \includegraphics[width=\linewidth]{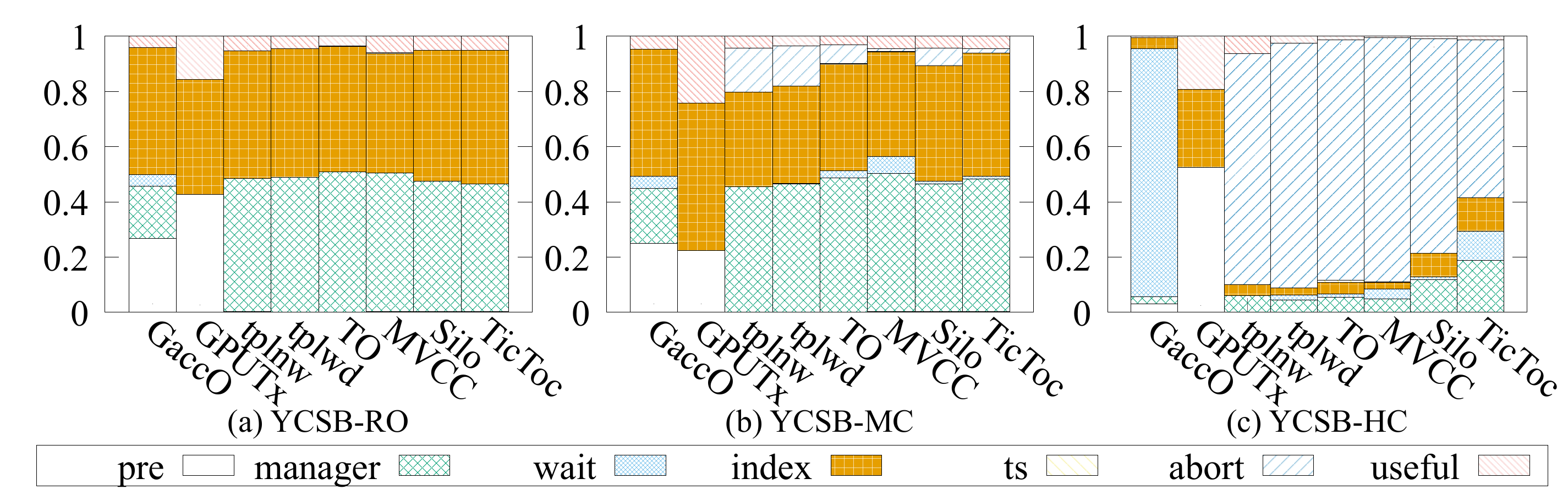}
    \vspace{-1em}
    \caption{Execution Time Breakdown. $wd=0,bs=32$}
    \label{fig:ycsb_bd}
  \end{minipage}
  \begin{minipage}[b]{0.495\textwidth}
      \centering
\includegraphics[width=\linewidth]{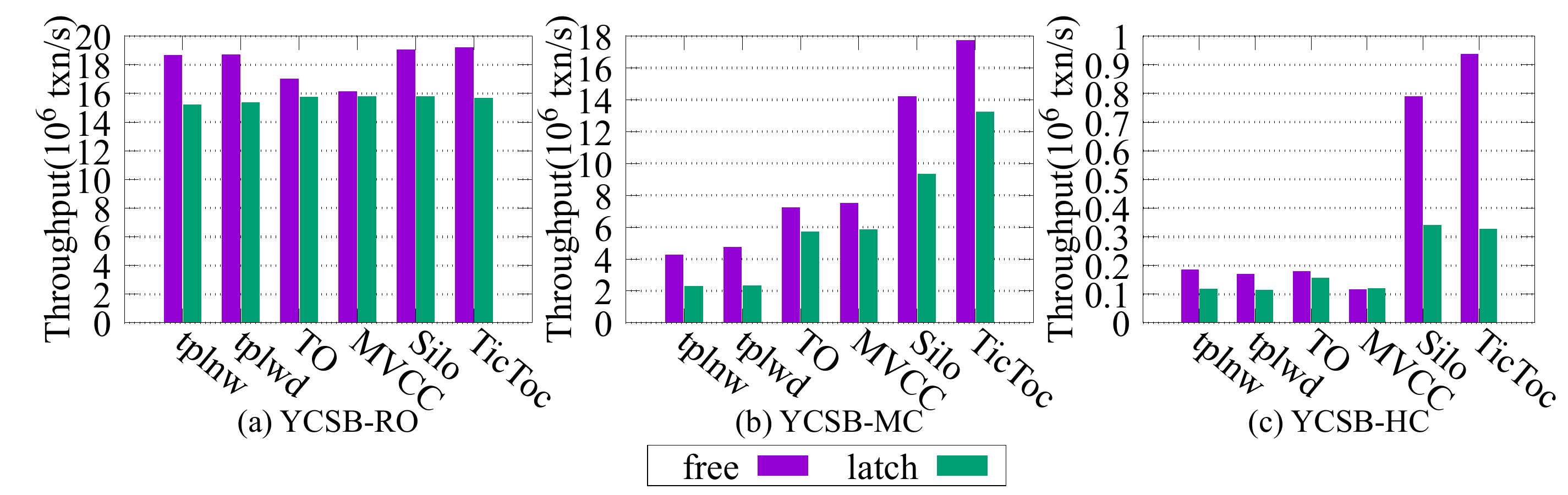}
      \caption{Latch-Free vs. Latch. $wd=0,bs=32$}
      \label{fig:ycsb_latch}
  \end{minipage}
\end{figure*}

\begin{figure}[t!]
  \centering
  \vspace{-1em}
  \includegraphics[width=\linewidth, height=0.13\textheight]{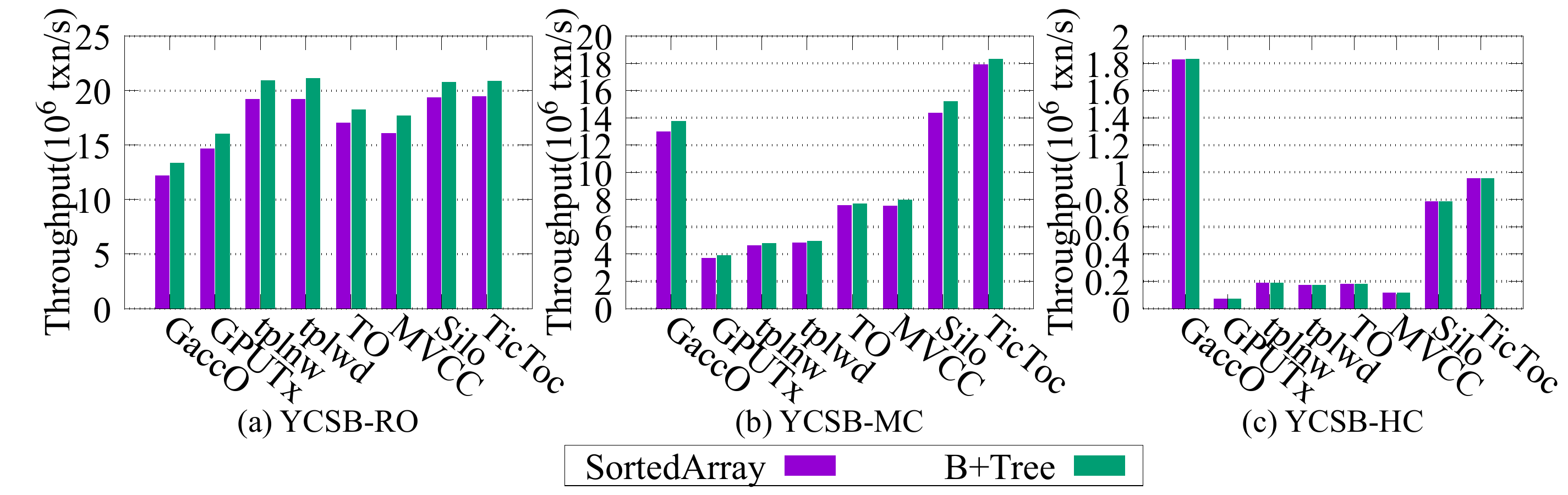}
  \vspace{-1.5em}
  \caption{SortedArray vs. B-Tree. $wd=0,bs=32$}
  \label{fig:ycsb_index}
\vspace{-2em}
\end{figure}



\stitle{\underline{Insight 2:}}
More worker threads per warp or more warps per block do not always lead to better performance. A higher $wd$ increases parallelism but also raises memory latency and the abort rate. While the exact relationship between $bs$ and performance is still unclear, the optimal $bs$ typically occurs around half of the maximum block size in low contention scenarios or at the maximum block size in high contention scenarios, providing a heuristic for selecting $bs$.


\subsection{Execution Time Breakdown}

\stitle{Exp-6: Execution Time Breakdown.}
To further identify performance bottlenecks, we measure the time spent in each stage of the transaction execution process. The duration of a stage is obtained by calculating the average time spent on that stage across all transactions in a batch.
In this experiment, schemes are tested with $wd=0,bs=32$ on the three preset YCSB benchmarks.

Figure~\ref{fig:ycsb_bd}(a) shows the execution time breakdown of CC schemes on YCSB-RO. The time consumption for the scheme managers, except for the conflict graph-based schemes, is similar, indicating that preprocessing is the main cause of performance lag in GPU-oriented schemes. Notably, although there is no need to wait, \textbf{GaccO} still incurs wait time because it treats both read and write operations as the same type of lock. Another key finding is that nearly one-third of the time is spent on index lookups (binary search), highlighting an area for optimization.

Figure~\ref{fig:ycsb_bd}(b) shows the execution time breakdown of CC schemes on YCSB-MC. The breakdown for \textbf{GaccO} on YCSB-MC is similar to that on YCSB-RO, primarily because it uses the same lock type for both read and write operations. For simplicity, we refer to the sum of the abort and wait proportions as the conflict-resolving proportion.

The order of conflict-resolving proportion between CC schemes is consistent with the order of throughput and abort in Figure~\ref{fig:ycsb_mc_diffbs},
which proves that conflict-resolving costs are the decisive factor in performance in this case.


Figure~\ref{fig:ycsb_bd}(c) shows the execution time breakdown of CC schemes on YCSB-HC.
We make the following observations:

\begin{enumerate}
    
\item  It is apparent that with the exception of \textbf{GPUTx}, the time consumption of waiting and aborting accounts for the majority of transaction execution time.

\item  It is abnormal that the aborting cost of \textbf{tpl\_wd} is greater than that of \textbf{tpl\_nw}, although the former spends more time waiting for locks.

\item  A similar situation happened with \textbf{MVCC}, it also spends more time on aborting than \textbf{TO} although it avoids rollbacks caused by read-write conflicts.

\item  For \textbf{TO} and \textbf{MVCC}, timestamp allocation takes only a tiny fraction of the time to be seen on the figures, both on YCSB-MC and YCSB-HC.

\item  OCC schemes continue to have the lowest conflict-resolving proportion even in a high contention environment.

\item  \textbf{TicToc} takes a more significant proportion of time waiting than \textbf{Silo} because \textbf{TicToc} spends time waiting for read timestamps to be updated.

\end{enumerate}

\stitle{\underline{Insight 3:}}
The key factor affecting the performance of CPU-oriented schemes on GPUs is the conflict-resolving overhead during contention. OCC schemes minimize this overhead by reducing the conflict-prone duration of transaction processing. They acquire locks only during the validation phase and release them quickly, whereas 2PL schemes acquire locks when writing for the first time and hold them throughout the transaction.

We observe that indexes account for a large proportion of the cost under YCSB-RO and YCSB-MC. We further evaluate the impact of different indexes, and the results are shown in Fig.~\ref{fig:ycsb_index}. It can be seen that replacing the sorted array with a B+ tree yields slight performance improvements under YCSB-RO and YCSB-MC, while the effect is negligible under YCSB-HC.

\subsection{Latch-Free vs. Latch}

\stitle{Exp-7: Latch-Free vs. Latch.}
All of our default CC scheme implementations adopt latch-free operations.
We want to know how much latch-free operations improve the performance.
Therefore, we re-implement all of the non-deterministic schemes using latches, test them with $wd=0,bs=32$ on three preset YCSB benchmarks, and compare their throughput with latch-free implementations.

We make the following observations based on Figure~\ref{fig:ycsb_latch}.

\begin{enumerate}
    
\item On YCSB-RO, the performance of each scheme's latching implementation is similar to and slightly lower than that of the latch-free version.

\item On YCSB-MC, the performance gap between the latch and latch-free versions of the same scheme has widened. The performance relationship between the latch versions of schemes is the same as the latch-free versions.

\item  On YCSB-HC, the performance gap between both versions of each scheme other than OCC has narrowed; even the latch version of \textbf{MVCC} outperforms the latch-free version.
\end{enumerate}

The latch-free operation involves more memory read/write operations (64 bits + tuple size) compared to a latch (32 bits) in a loop body. As contention increases, the greater number of loop executions amplifies the memory overhead, narrowing the performance gap. Additionally, the OCC read phase requires the latch version to read a row tuple, contributing to a significant performance gap.


\stitle{\underline{Insight 4:}} 
Effective optimization for CC schemes on GPUs must focus on reducing conflict-resolving overhead, which includes waiting time and the product of the average abort overhead and abort rate. Latch-free optimization reduces the cost of acquiring exclusive access to data, minimizing all conflict-resolving overhead factors. Since the waiting time of an aborted transaction counts towards rollback time, optimizations that reduce the abort rate by waiting may have limited impact, as seen with \textbf{tpl\_wd} and \textbf{MVCC} performance.


\subsection{TPC-C Results}

In the following, we test the \verb|Payment| and \verb|NewOrder| separately. 
For each transaction, we test the schemes with $wd=0, bs=32$.
Due to space limitations, we have placed the complete experiments conducted on TPC-C in our open-source repository ({\url{https://github.com/HKUSTDial/gCCTB}}).

\begin{figure}[t!]
  \centering
  \vspace{-3.5em}
  \setlength{\abovecaptionskip}{0pt}
  \subfloat[Total Throughput]{
  \includegraphics[width=0.49\linewidth]{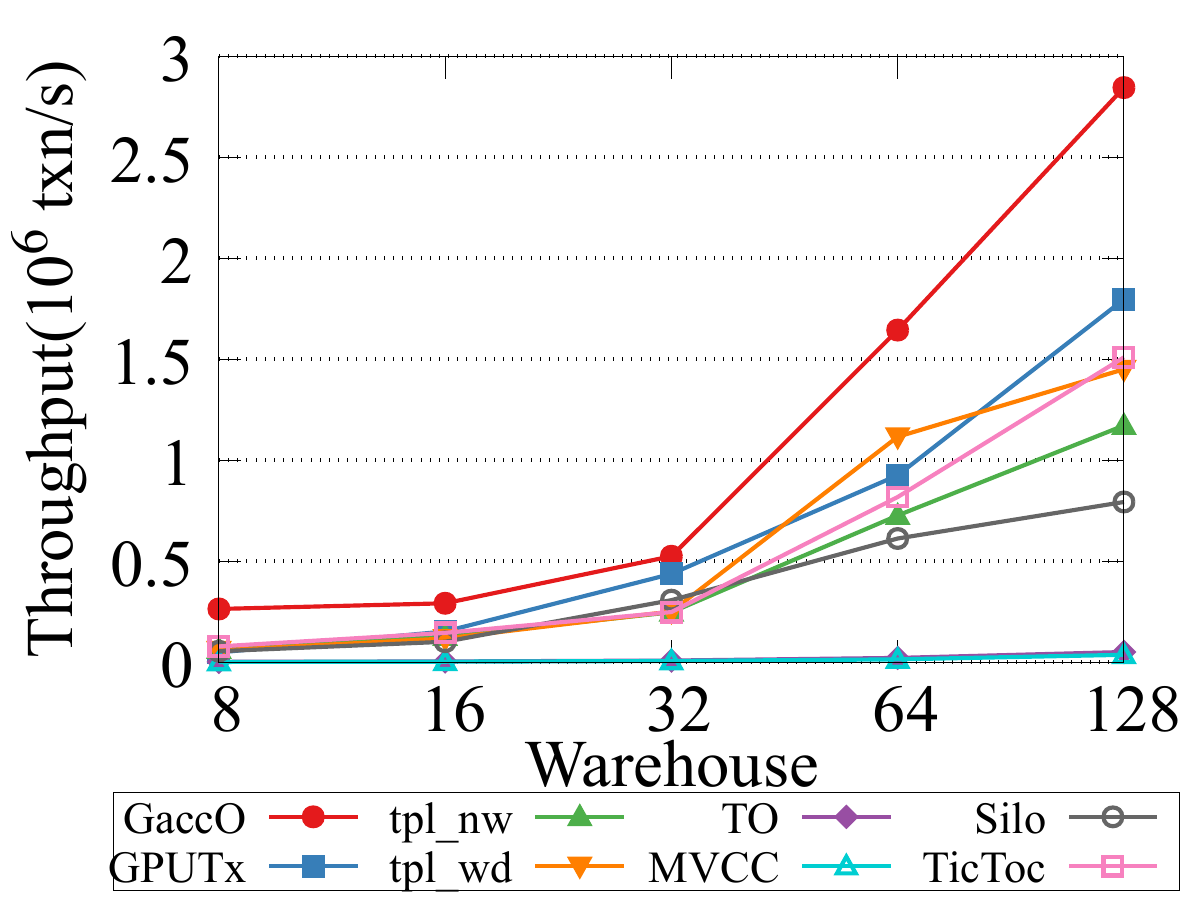}
  }
  \subfloat[Abort Rate]{
  \includegraphics[width=0.49\linewidth]{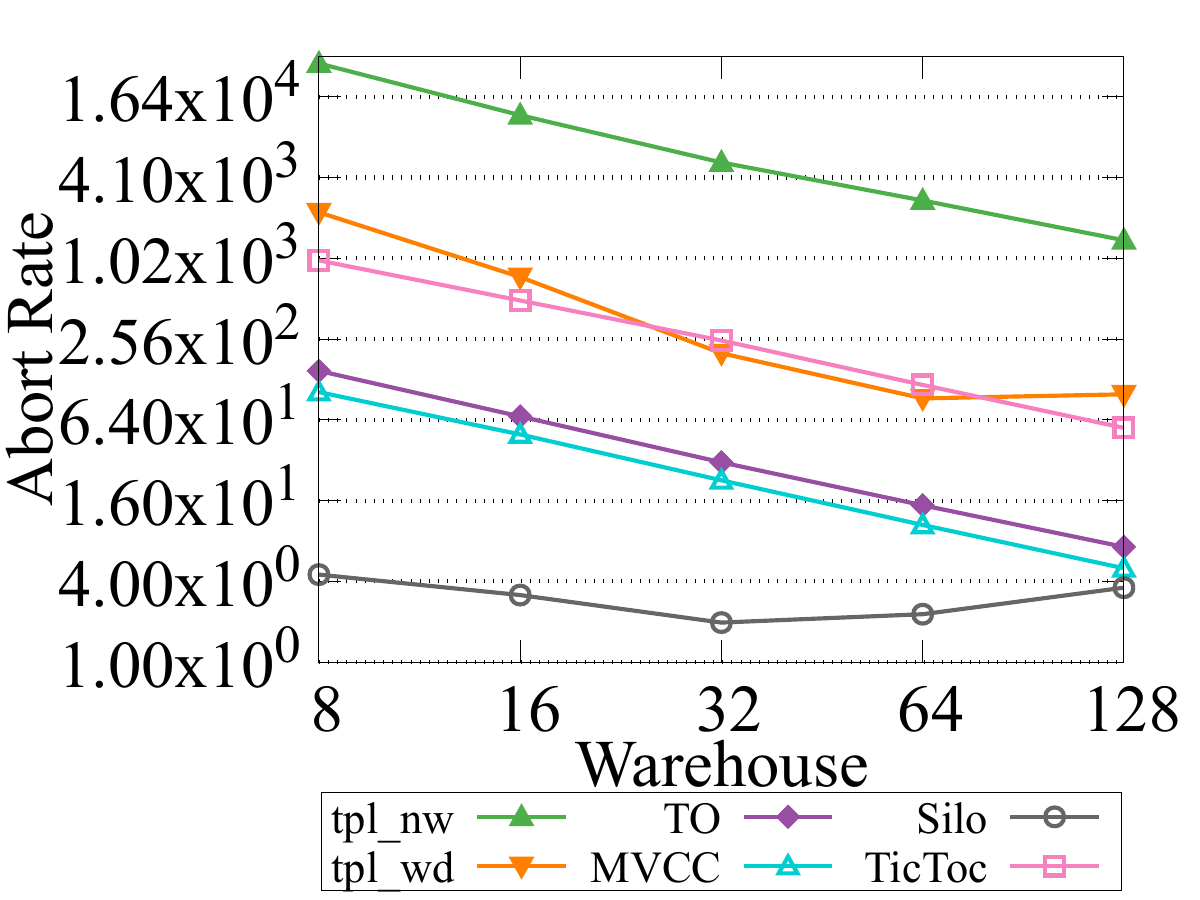}
  }
  \caption{TPC-C Payment. $wd=0, bs=32$}
  \label{fig:tpcc_pm_diffwh}
\vspace{-2em}
\end{figure}

\stitle{Exp-8: Performance on TPC-C Payment.}
Figure~\ref{fig:tpcc_pm_diffwh} shows the impact of warehouse number. Except for \textbf{mvcc}, the throughput of other schemes increases greatly with the increase of the number of warehouses.
The abort rates of \textbf{tpl\_wd} and \textbf{silo} go through a decrease and then an increase, with the lowest values appearing at 64 and 32 warehouses, respectively. 
The abort rates of other CC schemes decrease with the increase in the number of warehouses.


\begin{figure}[t!]
  \centering
  \vspace{-1em}
  \setlength{\abovecaptionskip}{0pt}
  \subfloat[Total Throughput]{
  \includegraphics[width=0.49\linewidth]{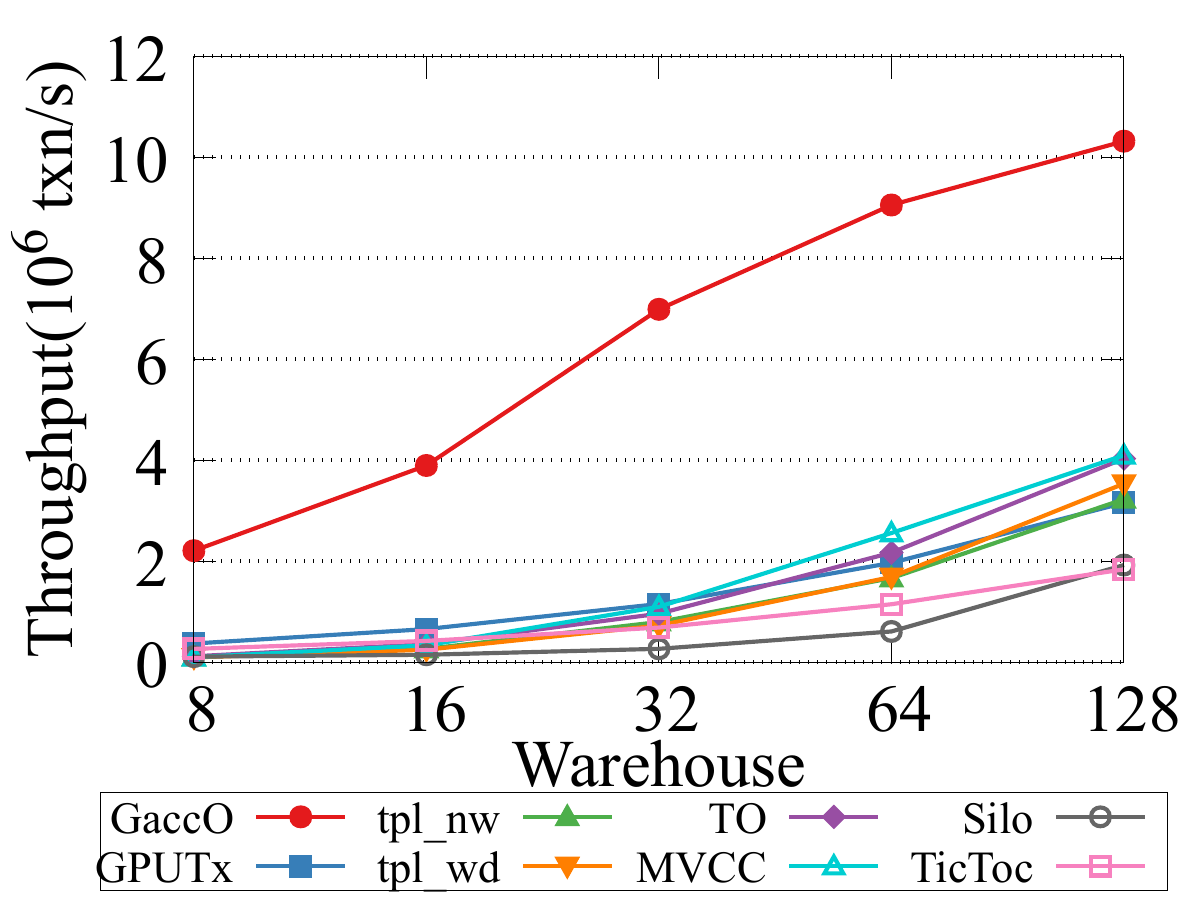}
  }
  \subfloat[Abort Rate]{
  \includegraphics[width=0.49\linewidth]{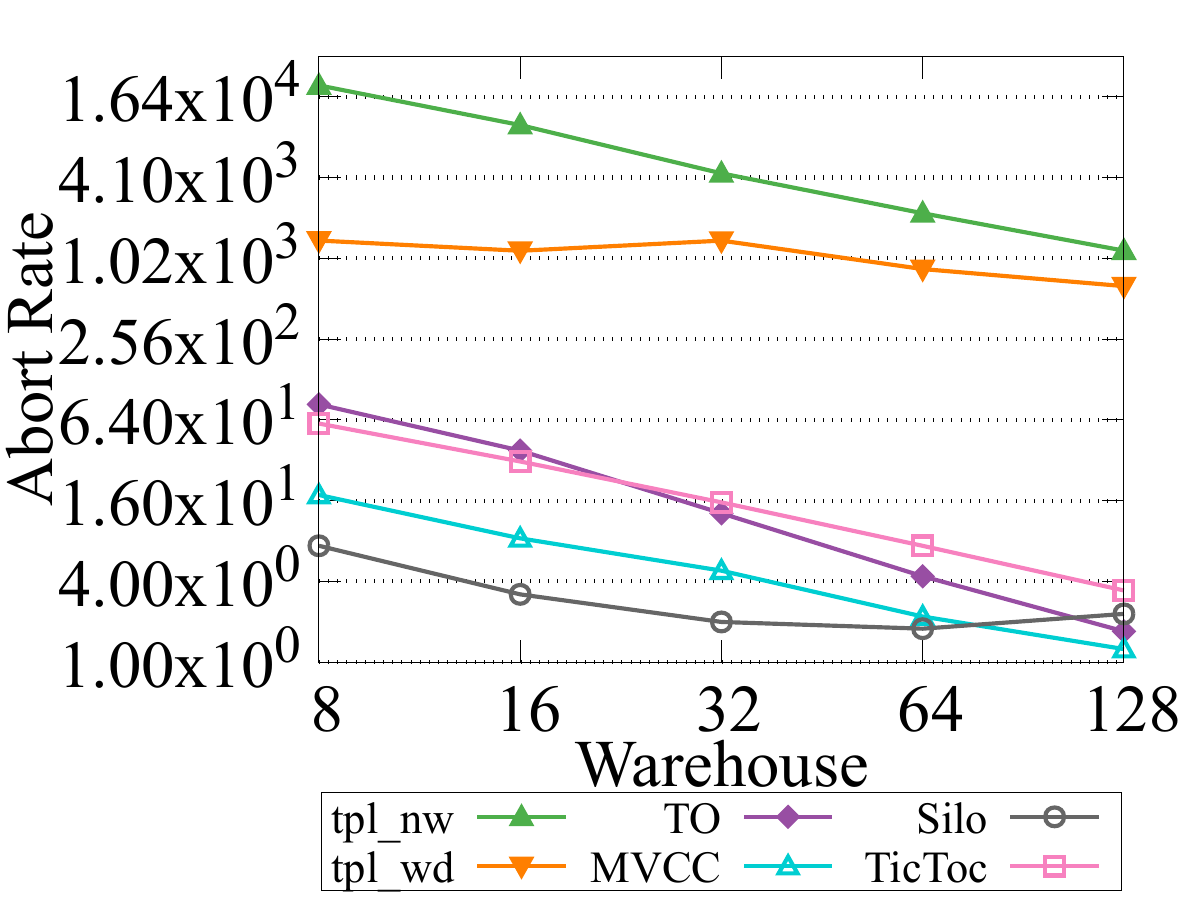}
  }
  \caption{TPC-C NewOrder $wd=0, bs=32$}
  \label{fig:tpcc_no_diffwh}
\vspace{-2em}
\end{figure}

\stitle{Exp-9: Performance on TPC-C NewOrder.}
Figure~\ref{fig:tpcc_no_diffwh} shows the impact of warehouse number.
The throughput of \textbf{gacco} is much higher than other schemes and even reaches $10^7$, while there is not much difference in the performance of other schemes,
The performance of all schemes increases monotonically with the increase in the number of warehouses. 
In terms of the abort rate, \textbf{silo} shows a trend of falling first and then rising.

\subsection{Main Findings}
\label{sec_findings}
We summarize the key findings as follows:

\stab \textit{(1)} When $\theta$ and $W$ are low ($\theta < 0.5, W=0.1$ or $W=0$ in our experiments), CPU-oriented schemes can outperform GPU-oriented schemes. However, \textbf{GaccO} has demonstrated an absolute advantage in high-conflict scenarios. When write operations occur, most of the CPU-oriented schemes face sharp performance degradation, while \textbf{GaccO} can keep performance stable.

\stab \textit{(2)} Among CPU-oriented schemes, performance follows the order: $\textbf{TicToc}>\textbf{Silo}>\textbf{MVCC} \approx \textbf{TO}>\textbf{tpl\_nw} \approx \textbf{tpl\_wd}$ when write operations are involved. This ranking aligns with the abortion rate.

\stab \textit{(3)} It is not that more worker threads in a warp are better. On YCSB-RO, the peak performance occurs around $wd=4$ because of the uncoalesced memory access problem. 
When the contention level increases, a smaller $wd$ reduces the intra-warp conflict and thus can improve the performance significantly.

\stab \textit{(4)} More warps in a block do not always improve performance. When contention is low, performance fluctuates sharply around $bs=16$ and reaches its peak, as shown by the vertical line in Figure~\ref{fig:ycsb_ro_heat}. This behavior is largely unaffected by $wd$, and though we don't yet fully understand the cause, it may serve as a heuristic for selecting an optimal $bs$.

\stab \textit{(5)} When contention is low, index overhead becomes a significant factor. In our current experiment, we use a sorted array with binary search. A well-designed GPU index could further accelerate transaction processing, complementing CC schemes.

\stab \textit{(6)} The decisive factor affecting the performance of CPU-oriented schemes on GPU is the conflict-resolving overhead when there exists contention. This is proven by the consistency of performance order, abort rate order, and conflict-resolving proportion order. 

\stab \textit{(7)}  An optimization that reduces the abort rate by waiting or multi-versioning may not perform well because they both 
increase the time spent on each attempt to execute a transaction.
Although searching on the version chain reduces the failures of read operations, it increases the memory access overhead.

\stab \textit{(8)} Latch-free strategy is a successful optimization, especially on OCC schemes when the contention level is high. It can reduce the cost of acquiring exclusive access to data items and, therefore, reduce all the factors of the conflict-resolving overhead.

Based on these findings, we provide several actionable guidelines for practitioners. When workloads are read-dominant and contention is low, CPU-oriented optimistic schemes (e.g., \textbf{Silo}, \textbf{TicToc}) should be preferred. In contrast, under high write intensity and heavy contention, GPU-oriented schemes such as \textbf{GaccO} are more suitable. Regarding GPU kernel parameters, we recommend using a smaller warp density ($wd \le 2$) in high-contention settings to reduce aborts, while choosing a larger block size ($bs$ close to the maximum) to maximize throughput. For low-contention workloads, setting $bs$ around half of the maximum value often yields the best performance. Finally, latch-free implementations are strongly advised, as they substantially reduce conflict resolution overhead compared to latch-based counterparts.

\section{Related Work}
\label{sec_rel}
\vspace{-.25em}

\stitle{GPU-accelerated DBMS.}
Several GPU-accelerated DBMS~\cite{bress2014design,root2016mapd,lee2021art,heavydb} have been developed for both research and commercial purposes, primarily to accelerate analytical processing. Key database operations such as sorting~\cite{satish2009designing,satish2010fast,stehle2017memory,maltenberger2022evaluating}, join~\cite{rui2015join,sioulas2019hardware,lutz2020pump,rui2020efficient,lutz2022triton}, and compaction~\cite{xu2020luda} now have GPU-optimized versions. Recent trends focus on using high-speed interconnects to harness the power of multiple GPUs. Additionally, GPU acceleration is also applied to indexing~\cite{awad2019engineering,awad2022gpu,zhang2015mega}.

\stitle{OLTP Benchmarking.}
DBx1000~\cite{staring14} is a lightweight, main-memory DBMS designed to assess the scalability of CC schemes. It implements seven CC schemes and simulates scaling up to 1024 cores. Initially developed to study performance variations with increasing cores, it has become a testbed for evaluating new CC schemes.
CCBench~\cite{tanabe2020ccbench} is another main-memory DBMS for benchmarking CC schemes. It supports seven schemes, seven optimization methods, and seven workload parameters, with an emphasis on fairness in comparisons. CCBench ensures fairness by providing shared core modules, such as access methods and thread-local data structures, used across all schemes.
However, both DBx1000 and CCBench are CPU-based and do not support GPUs, motivating the creation of a new testbed specifically for benchmarking CC schemes on GPUs.


\stitle{In-Memory Transaction Processing.}
We leave the implementation and benchmarking of schemes combining MVCC and OCC for future work. These schemes are used in in-memory transaction processing engines such as Hekaton~\cite{diaconu2013hekaton,larson2011high} and Cicada~\cite{lim2017cicada}. Additionally, OCC can be combined with pessimistic locking, as seen in Mostly-Optimistic Concurrency Control (MOCC)~\cite{mocc16}, which integrates pessimistic read locks to prevent writer-reader conflicts. Another hybrid scheme, Pessimistic Locking and Optimistic Reading (PLOR)~\cite{plor22}, offers high throughput and low tail latency.

\section{Conclusion} \label{sec_conclu}
\vspace{-.25em}

In this paper, we presented \ourmethod{}, the first testbed designed to evaluate concurrency control schemes on GPUs. We implemented and evaluated eight CC schemes, including six well-established CPU-oriented schemes and two designed for GPUs. Our comprehensive evaluations using YCSB and TPC-C benchmarks and various GPU-specific parameters yielded several key insights. These findings suggest that lessons from CPU-designed schemes can inform the development of more effective CC schemes tailored for GPU environments.



\section{Acknowledgements}

This work was supported by the NSF of China (62402409), Youth S\&T Talent Support Programme of GDSTA (SKXRC2025461), and Young Talent Support Project of Guangzhou Association for Science and Technology (QT-2025-001).  

\section{AI-Generated Content Acknowledgement}

During the preparation of this work, the authors used ChatGPT and Gemini in order to improve the readability and language quality of the manuscript. These tools were employed specifically to refine the writing style and correct grammatical errors throughout the text. After using these tools, the authors reviewed and edited the content as needed and take full responsibility for the content of the publication.

\balance

\bibliographystyle{ieeetr}
\bibliography{refs/ref}


\end{document}